\documentclass[12pt]{article}
\usepackage{epsfig}
\begin{document}
\title{On the possibility of testing the two-peak structure of the LHCb hidden-charm strange pentaquark
$P_{cs}(4459)^0$ in near-threshold antikaon-induced charmonium production on protons and nuclei}
\author{E. Ya. Paryev \\
{\it Institute for Nuclear Research of the Russian Academy of Sciences}\\
{\it Moscow, Russia}}

\renewcommand{\today}{}
\maketitle

\begin{abstract}
Accounting for the LHCb observation that the reported hidden-charm strange pentaquark $P_{cs}(4459)^0$
can split into two substructures, $P_{cs}(4455)^0$ and $P_{cs}(4468)^0$, with a mass difference of 13 MeV
as well as the newly observed hidden-charm pentaquark resonance $P_{cs}(4338)^0$ with strangeness,
we study within the double-peak scenario for the $P_{cs}(4459)^0$ state
the near-threshold $J/\psi$ meson production from protons and nuclei
by considering incoherent direct non-resonant (${K^-}p \to {J/\psi}\Lambda$) and two-step resonant
(${K^-}p \to P_{csi}^0 \to {J/\psi}\Lambda$, $i=1$, 2, 3; $P_{cs1}^{0}=P_{cs}(4338)^0$,
$P_{cs2}^{0}=P_{cs}(4455)^0$, $P_{cs3}^{0}=P_{cs}(4468)^0$) charmonium production processes
with the main goal of clarifying the possibility to observe within this scenario both above two substructures
contributing to the $P_{cs}(4459)^0$ state and the $P_{cs}(4338)^0$ resonance in this production.
We calculate the absolute excitation functions, energy and momentum distributions
for the non-resonant, resonant and for the combined (non-resonant plus resonant) production
of $J/\psi$ mesons on protons as well as
on carbon and tungsten target nuclei at near-threshold incident antikaon beam momenta by assuming
the spin-parity assignments of the hidden-charm resonances
$P_{cs}(4338)^0$, $P_{cs}(4455)^0$ and $P_{cs}(4468)^0$ as $J^P=(1/2)^-$, $J^P=(1/2)^-$
and $J^P=(3/2)^-$ within four different realistic choices for the branching ratios
of their decays to the ${J/\psi}\Lambda$ mode (0.125, 0.25, 0.5 and 1\%) as well as for two options
for the branching fraction of their decays to the $K^-p$ channel (0.01 and 0.001\%).
We show that these combined observables reveal clear sensitivity to these scenarios.
Hence, they may be an important tool to provide further evidence for the existence of
the above strange hidden-charm pentaquark resonances.
\end{abstract}

\newpage

\section*{1. Introduction}
The study of pentaquark states, comprising four quarks and an antiquark, has received considerable
interest in recent years. The first observation of the hidden-charm pentaquark resonances
$P_c(4380)^+$ and $P_c(4450)^+$ in the ${J/\psi}p$ invariant mass spectrum of the
$\Lambda^0_b \to K^-({J/\psi}p)$ decays has been reported by the LHCb experiment [1]
\footnote{$^)$For better readability of this paper, the old exotic hadron naming scheme is used throughout
it instead of the new one proposed in Ref. [2]. According to the latter one, the $P_{\psi}^N$ would be the new name for the $P_c$ states.}$^)$
.
With the inclusion
of additional data, it was found that the $P_c(4450)^+$ state is composed of two narrow overlapping peaks,
$P_c(4440)^+$ and $P_c(4457)^+$. In addition, a new narrow state, the $P_c(4312)^+$ was discovered [3]. Recently, evidence for
a new narrow hidden-charm pentaquark denoted as $P_{c}(4337)^+$ was found in the invariant mass spectrum of the ${J/\psi}p$ in the $B_s^0 \to {J/\psi}p{\bar p}$ decays [4].
The minimal quark structure of the above pentaquarks is $|P^+_c>=|uudc{\bar c}>$.
In a molecular scenario, due to the closeness of the observed
$P_c(4312)^+$ and $P_c(4440)^+$, $P_c(4457)^+$ masses to the ${\Sigma^+_c}{\bar D}^0$ and
${\Sigma^+_c}{\bar D}^{*0}$ thresholds, the $P_c(4312)^+$
resonance can be, in particular, interpreted as an S-wave ${\Sigma^+_c}{\bar D}^0$ bound state, while the
$P_c(4440)^+$ and $P_c(4457)^+$ as S-wave ${\Sigma^+_c}{\bar D}^{*0}$ bound molecular states [5--16], though
their inner structure (charmed baryon--charmed antimeson two-body molecule or compact pentaquark state)
is still not determined and there are alternative explanations too [17--20]. Due to lacking nearby meson-baryon
thresholds, the $P_c(4337)^+$ is difficult to accommodate within the molecular picture [21, 22]. But, it is more
natural to view it as the compact pentaquark state [19, 23].
The LHCb Collaboration announced recently also the discovery of two new narrow hidden-charm strange
pentaquark states, $P_{cs}(4459)^0$ [24] and $P_{cs}(4338)^0$ [25], with minimal quark content $|udsc{\bar c}>$
in the invariant mass spectra of the ${J/\psi}\Lambda$ in the $\Xi^-_b \to K^-({J/\psi}\Lambda)$ and $B^- \to {\bar p}({J/\psi}\Lambda)$ decays
\footnote{$^)$Following the new naming convention in [2], in Ref. [25] these newly observed states are named as $P_{{\psi}s}^{\Lambda}(4459)^0$ and $P_{{\psi}s}^{\Lambda}(4338)^0$.}$^)$
.
The Breit-Wigner masses and total widths of these exotic states were measured to be [24, 25]:
\begin{equation}
P_{cs}(4459)^0:\,\,\,M=4458.8\pm2.9^{+4.7}_{-1.1}~{\rm MeV},\,\,\,\Gamma=17.3\pm6.5^{+8.0}_{-5.7}~{\rm MeV};
\end{equation}
\begin{equation}
P_{cs}(4338)^0:\,\,\,M=4338.2\pm0.7\pm0.4~{\rm MeV},\,\,\,\Gamma=7.0\pm1.2\pm1.3~{\rm MeV}.
\end{equation}
While the spin-parity quantum numbers $J^P$ of the $P_{cs}(4459)^0$ resonance were not experimentally determined
due to a limited signal yield, the spin of the $P_{cs}(4338)^0$ pentaquark candidate is determined to be $J=1/2$
and negative parity is preferred [25]. Since the new strange $P_{cs}(4459)^0$ and $P_{cs}(4338)^0$ pentaquarks are found
close to thresholds for the production of ordinary baryon-meson states $\Xi_c{\bar D}^*$ and $\Xi_c{\bar D}$
\footnote{$^)$The pole mass of the  $P_{cs}(4459)^0$ state is just about 19 MeV below the
$\Xi_c^0{\bar D}^{*0}$ threshold [24]. The central value of the $P_{cs}(4338)^0$ mass is only 0.8 and 2.9 MeV
above, respectively, the $\Xi_c^+D^{-}$ and $\Xi_c^0{\bar D}^{0}$ thresholds [26].}$^)$
,
it is natural to interpret their as the hidden-charm singly-strange $\Xi_c{\bar D}^*$ and $\Xi_c{\bar D}$ molecules.
Such plausible molecular interpretation of the $P_{cs}(4459)^0$ and $P_{cs}(4338)^0$ exotic states with strangeness
was supported by many current theoretical studies (see, e.g., Refs. [23--38]) and is commonly accepted.
However, there exists another explanations of the $P_{cs}$ states, based both on the compact quark model
[19, 39--42] and on other molecular picture [43], in which it was found that due to the extra attraction
obtained in the $\Xi_c{\bar D}^*$, $\Xi_c{\bar D}$ pairs the $P_{cs}(4459)^0$ and $P_{cs}(4338)^0$ states
are associated, respectively, to the $\Xi_c^{\prime}{\bar D}$ and $\Xi_c{\bar D}^*$.
It should be pointed out that the existence of hidden-charm strange pentaquark resonances $P_{cs}$ has been
predicted before the LHCb experiment [24] in some earlier papers (see, for example, Refs. [15, 44--53]).
It is also worth noting that the large $Q$-values of the decays $P_{cs}(4459)^0 \to {J/\psi}\Lambda$ and
$P_{cs}(4338)^0 \to {J/\psi}\Lambda$ (in which the pentaquarks $P_{cs}(4459)^0$ and $P_{cs}(4338)^0$ were
observed) -- respectively, 246 MeV and 126 MeV -- and unnaturally small for such
$Q$-values widths of the $P_{cs}(4459)^0$ and $P_{cs}(4338)^0$ states about of 17 MeV and 7 MeV tell
us in favor of $\Xi_c{\bar D}^*$ and $\Xi_c{\bar D}$ molecular interpretation of these states since in this interpretation their decay into the ${J/\psi}\Lambda$ is naturally suppressed due to the fact that
two $c$ and ${\bar c}$ quarks, belonging, respectively, to the $\Xi_c$ baryon and ${\bar D}^*$, ${\bar D}$ mesons,
in hadronic molecules are well-separated and it is unlikely for them to be close to each other
to form a single $J/\psi$ meson, which provides decay-suppressing mechanism [26, 29].
Along the way of molecular scenario for the $P_{cs}(4459)^0$ resonance, one can easily find its possible spin-parity
quantum numbers. Considering it as pure $\Xi_c{\bar D}^*$ molecule, in which $\Xi_c$ baryon with $J^P=(1/2)^+$
and antimeson ${\bar D}^*$ with $J^P=(1)^-$ are in a relative S-wave, we get that the spin-parity of the
$P_{cs}(4459)^0$ is either $J^P=(1/2)^-$ or $J^P=(3/2)^-$ (cf. Ref. [26])
\footnote{$^)$Similarly,  identifying the $P_{cs}(4338)^0$ state as the S-wave $\Xi_c(1/2)^+{\bar D}(0^-)$
molecular state, we obtain that its spin-parity combination would be $J^P=(1/2)^-$ -- what exactly was observed
in the LHCb experiment [25].}$^)$
.
This conclusion was supported by more elaborate theoretical approaches (see, e.g., Refs. [19, 30, 34, 36, 37, 39, 40, 41, 48, 54]), in which the $P_{cs}(4459)^0$ is considered as $\Xi_c{\bar D}^*$ molecular
pentaquark state [30, 34, 36, 37, 48, 54] or as compact pentaquark state [19, 39, 40, 41].
In the absence of coupled-channel effects the heavy-quark spin symmetry predicts the two $\Xi_c{\bar D}^*$
states (with total spin $J=1/2$ and $J=3/2$, respectively) to be degenerate [55]. Their inclusion breaks the
spin degeneracy of the $P_{cs}(4459)^0$ and generates a hyperfine splitting between the two
spin configurations [21, 27, 31, 32, 33, 35, 38, 50, 55, 56], in which the $J^P=(1/2)^-$ $\Xi_c{\bar D}^*$ pentaquark
will be lighter than the $J^P=(3/2)^-$ one [27, 38, 50] and vice versa [21, 31, 32, 33, 35, 55, 56].
This means that in the mass region of the $P_{cs}(4459)^0$ (slightly below the $\Xi_c{\bar D}^*$ threshold)
may exist two almost degenerate narrow overlapping
$\Xi_c{\bar D}^*$ molecular states, similar to the case for the non-strange hidden-charm pentaquark $P_{c}(4450)^+$, which can be equally replaced by two pentaquark substructures $P_{c}(4440)^+$ and $P_{c}(4457)^+$ [3].
In particular, in this region Ref. [50] predicts, within the chiral effective field theory, the masses of the two
$\Xi_c{\bar D}^*$ molecular states to be:
$$
\Xi_c{\bar D}^*~{\rm with}~J^P=(1/2)^-:\,\,\,M=4456.9^{+3.2}_{-3.3}~{\rm MeV},
$$
\begin{equation}
\Xi_c{\bar D}^*~{\rm with}~J^P=(3/2)^-:\,\,\,M=4463.0^{+2.8}_{-3.0}~{\rm MeV}.
\end{equation}
Motivated by these findings, the LHCb Collaboration tested the hypothesis of the two-peak structure of the
$P_{cs}(4459)^0$ resonance [24] with the predicted in [50] $J^P$ values.
It was found [24] that this resonance can also well be described by two pentaquark resonances
with mass difference of 13 MeV, named in the present paper as $P_{cs}(4455)^0$ and $P_{cs}(4468)^0$ (cf. [21, 26]).
Their masses and widths were determined to be:
$$
P_{cs}(4455)^0:\,\,\,M=4454.9\pm2.7~{\rm MeV},\,\,\,\Gamma=7.5\pm9.7~{\rm MeV};
$$
\begin{equation}
P_{cs}(4468)^0:\,\,\,M=4467.8\pm3.7~{\rm MeV},\,\,\,\Gamma=5.2\pm5.3~{\rm MeV}.
\end{equation}
We see that the 'experimental' masses (4) are in excellent agreement with the calculated ones of Eq. (3).
This fact and the theoretical predictions of the double-peak structure of the $\Xi_c{\bar D}^*$ molecule
in the energy region of the $P_{cs}(4459)^0$ resonance, discussed above,
make us suggest that the $P_{cs}(4459)^0$ should indeed contain two substructures.
And from now on, we will assume the existence of the two $P_{cs}(4455)^0$ and $P_{cs}(4468)^0$ peaks
in spite of the fact that the LHCb analysis of the current data sample cannot confirm or disprove the
two-peak solution (4). One may hope that the future more precise LHCb measurements and other ones,
stimulated, in particular, by the results of the present work, would indeed shed light
on whether the double-peak interpretation of the $P_{cs}(4459)^0$ state is correct.

Remarkably, another hidden-charm exotic state with open strangeness -- the strange hidden-charm
tetraquark state $Z_{cs}(3985)^-$, decaying into $D_s^-D^{*0}$ and $D_s^{*-}D^0$,
has been observed in 2020 by the BESIII Collaboration in the processes
$e^+e^- \to K^+(D_s^-D^{*0}+D_s^{*-}D^0)$ [57]. This state has the minimal quark content
$Z_{cs}(3985)^-=|{\bar u}sc{\bar c}>$. It is in the proximity of the $D_s^-D^{*0}/D_s^{*-}D^0$
mass thresholds and can be interpreted [58, 59] as strange molecular partner
with hadronic molecular configuration $D_s^-D^{*0}-D_s^{*-}D^0$
of the non-strange tetraquark resonance $Z_{c}(3900)^-$, having the valence quark content
$Z_{c}(3900)^-=|{\bar u}dc{\bar c}>$ and possible hadronic molecular structure $D^-D^{*0}-D^{*-}D^0$.
Moreover, in 2022 year the BESIII Collaboration reported [60] evidence for the neutral $Z_{cs}(3985)^0$
state in the processes $e^+e^- \to K^0_S(D_s^+D^{*-}+D_s^{*+}D^-)$, whose mass and width are close to those of the
charged $Z_{cs}(3985)^-$. Hence, the $Z_{cs}(3985)^0$ can be regarded as the isospin partner
of the $Z_{cs}(3985)^-$. Earlier, the LHCb Collaboration reported the first observation of exotic states
$Z_{cs}(4000)^+$ and $Z_{cs}(4220)^+$ with a new quark content $|u{\bar s}c{\bar c}>$ in the mass spectra of
the ${J/\psi}K^+$  in $B^+ \to {J/\psi}{\phi}K^+$ decays [61]. Furthermore, the first results of searching
for a heavier partner of the observed $Z_{cs}(3985)^-$ state, denoted as $Z_{cs}^{{\prime}-}$, in the process
$e^+e^- \to K^+D_s^{*-}D^{*0}$ (c.c.) with the BESIII detector have been reported in very recent publication [62].
In addition, one needs to note that possible hidden-charm pentaquarks, having two and three strange quarks in their
valence composition (or pentaquarks with double and triple strangeness), have been theoretically investigated
in Refs. [19, 49, 63, 64, 65, 66]. They are at present an experimentally virgin territory.
But one may hope that  they will be discovered in the foreseeable future at the LHC as well.
It should be noticed that the investigation of magnetic moments of the hidden-charm pentaquark states without
strangeness, with strangeness and with double strangeness, which play an important role in understanding their
internal structure, has been performed, in particular, in Ref. [67] within the QCD light-cone sum rules.
Extensive reviews of the recent experimental and theoretical progresses in the field of heavy hadronic molecular states are presented in Refs. [18, 68].

Confirming the $P_{cs}(4459)^0$ and $P_{cs}(4338)^0$ states, observed in the $\Xi^-_b$ and $B^-$ decays at LHCb,
in other processes would provide independent and complementary verification of their existence as well as
would be helpful both to understand their nature and to pin down their resonance parameters.
In that way, the production of
the single-peak $P_{cs}(4459)^0$ pentaquark in antikaon-induced and photon-induced reactions on nuclear targets
was discussed in Refs. [69, 70] and [71], correspondingly.
In fact, in Ref. [69] the contribution of this pentaquark to the elementary $K^-p \to {J/\psi}\Lambda$ reaction
has been determined, employing two different theoretical approaches, i.e., the effective Lagrangian and the
Regge models. It was shown that the $P_{cs}(4459)^0$ can also be searched for through
a scan of the total and differential cross sections of this reaction. In our work [70], we have calculated
the absolute excitation functions, energy and momentum distributions
for the direct non-resonant ($K^-p \to {J/\psi}\Lambda$), two-step resonant
($K^-p \to P_{cs}(4459)^0 \to {J/\psi}\Lambda$) and for the combined (non-resonant plus resonant) production
of $J/\psi$ mesons off protons as well as off carbon and tungsten target nuclei at near-threshold
incident antikaon energies by assuming the spin-parity assignment of the $P_{cs}(4459)^0$ resonance
as $J^P=(3/2)^-$ within six different scenarios for the branching ratio $Br[P_{cs}(4459)^0 \to {J/\psi}\Lambda]$
of the decay $P_{cs}(4459)^0 \to {J/\psi}\Lambda$, namely:
$Br[P_{cs}(4459)^0 \to {J/\psi}\Lambda]=$ 1, 3, 5, 10, 15 and 50\%.
We also have shown that the combined observables considered reveal definite sensitivity to these scenarios, which
means that they may be an important tool to provide further evidence for the existence of the pentaquark
$P_{cs}(4459)^0$ resonance and to get valuable information on its decay rate to the ${J/\psi}\Lambda$ final state.
It should be noted that the energy of the negative kaon beam, which will be available at the K10 beam line in the extended J-PARC Hadron Experimental Facility [72], will be sufficient to observe the $P_{cs}(4459)^0$ pentaquark
via the $K^-p \to {J/\psi}\Lambda$ process. The possibility of searching for this pentaquark in
${\gamma}p \to K^+P_{cs}(4459)^0$ reaction, assuming that it with the spin-parity combinations $J^P=(1/2)^-$ or $J^P=(3/2)^-$ can be interpreted as a $\Xi_c{\bar D}^*$ molecule or as a compact pentaquark, has been investigated
in Ref. [71]. It was shown that the total cross section of this reaction, calculated in the latter picture of the
$P_{cs}(4459)^0$, is quite different from that, obtained in molecular scenario of the $P_{cs}(4459)^0$, which can be
used to test the nature of the $P_{cs}(4459)^0$ state [71]. And finally, in the recent work [73], based on the
hadronic molecular picture, the authors have estimated the semi-inclusive electroproduction rates, in particular, of the $P_{cs}(4459)^0$ and $P_{cs}(4338)^0$ multiquark states at the proposed electron-ion colliders in China (EicC),
US (EIC) and at the proposed 24 GeV upgrade of CEBAF.

 In the present article, assuming in view of the aforementioned that the $P_{cs}(4459)^0$ resonance splits
into two peaks, $P_{cs}(4455)^0$ and $P_{cs}(4468)^0$, with $J^P=(1/2)^-$ and $J^P=(3/2)^-$, respectively,
we consider the contribution both from these peaks and from the $P_{cs}(4338)^0$ state to near-threshold
$J/\psi$ meson production by antikaons on protons and nuclei by adopting
the standard Breit-Wigner shape for this contribution and by employing an available at present
information on the total and differential cross sections of the direct $K^-p \to {J/\psi}\Lambda$
process to estimate the background contribution. The main goal of this consideration is also to clarify the
possibility of experimental observation of the above two peaks and the $P_{cs}(4338)^0$ state in this production.
The consideration is strictly based on the model, developed in Ref. [70].
We briefly recall its main assumptions and describe, where
necessary, the corresponding extensions. We present the predictions obtained within this
extended model for the $J/\psi$ excitation functions, energy and momentum distributions
in ${K^-}p$ as well as in ${K^-}$$^{12}$C and ${K^-}$$^{184}$W reactions at
near-threshold incident antikaon momenta assuming different scenarios for the branching fractions
of the decays $P_{cs}(4338/4455/4468)^0 \to {K^-}p$ and $P_{cs}(4338/4455/4468)^0 \to {J/\psi}\Lambda$.
These predictions may serve as guidance for future dedicated experiment
at the J-PARC facility devoted to the study of possible two-peak structure of the $P_{cs}(4459)^0$ resonance.

\section*{2. Theoretical framework}

\subsection*{2.1. Direct non-resonant $J/\psi$ production process}

Direct non-resonant charmonium production on nuclear targets in the near-threshold laboratory incident
$K^-$ beam momentum region $8.844~{\rm GeV/c} \le p_{K^-} \le 10.200~{\rm GeV/c}$
\footnote{$^)$In which the pentaquark states $P_{cs}(4338)^0$, $P_{cs}(4455)^0$ and
$P_{cs}(4468)^0$ are concentrated and where they can be observed in the $K^-p$ reactions (see Figs. 1 and 2 below).
We recall that the threshold (resonant) momenta $p_{K^-}^{\rm R1}$,
$p^{\rm R2}_{K^-}$ and $p^{\rm R3}_{K^-}$ for the production of the $P_{cs}(4338)^0$,
$P_{cs}(4455)^0$ and $P_{cs}(4468)^0$ resonances with central values of their masses
$M_{cs1}=4338.2$ MeV, $M_{cs2}=4454.9$ MeV and $M_{cs3}=4467.8$ MeV (cf. Eqs. (2), (4))
on a free target proton being at rest are $p^{\rm R1}_{K^-}=9.417$ GeV/c, $p^{\rm R2}_{K^-}=9.965$ GeV/c
and $p^{\rm R3}_{K^-}=10.026$ GeV/c, respectively.}$^)$
may proceed via the following elementary process, which has the lowest free ${J/\psi}\Lambda$ production
threshold momentum (8.844 GeV/c) and in which the strangeness $S=-1$ is conserved [69, 70]:
\begin{equation}
K^-+p \to J/\psi+\Lambda.
\end{equation}
Before going further, let us get a feeling about kinematic characteristics of $J/\psi$ mesons -- laboratory
polar production angles and momenta (total energies), allowed in this process in the simpler case of a
free target proton being at rest at an incident resonant antikaon momenta of 9.417, 9.965 and 10.026 GeV/c
of our main interest. The kinematics of two-body reaction with a threshold
(as in our present case) indicate that the laboratory polar $J/\psi$ production angle $\theta_{J/\psi}$
varies from 0 to a maximal value $\theta^{\rm max}_{J/\psi}$, i.e.:
\begin{equation}
     0 \le \theta_{J/\psi} \le \theta^{\rm max}_{J/\psi},
\end{equation}
where
\begin{equation}
 \theta^{\rm max}_{J/\psi}={\rm arcsin}[(\sqrt{s}p^{*}_{J/\psi})/(m_{J/\psi}p_{K^-})].
\end{equation}
Here, the $J/\psi$ c.m. momentum $p^*_{J/\psi}$ is determined by the equation
\begin{equation}
p_{J/\psi}^*=\frac{1}{2\sqrt{s}}\lambda(s,m_{J/\psi}^{2},m_{\Lambda}^2),
\end{equation}
in which the quantities $m_{J/\psi}$ and $m_{\Lambda}$ are the free space charmonium and
$\Lambda$ hyperon masses and
the vacuum collision energy squared $s$ and function $\lambda(x,y,z)$
are defined, respectively, by the formulas
\begin{equation}
  s=(E_{K^-}+m_p)^2-{p}_{K^-}^2,\,\,\,\,E_{K^-}=\sqrt{m^2_{K^-}+p^{2}_{K^-}},
\end{equation}
\begin{equation}
\lambda(x,y,z)=\sqrt{{\left[x-({\sqrt{y}}+{\sqrt{z}})^2\right]}{\left[x-
({\sqrt{y}}-{\sqrt{z}})^2\right]}}.
\end{equation}
The quantities $m_{p}$ and $m_{K^-}$, entering into Eq. (9), denote the vacuum proton and
$K^-$ meson masses. From Eq. (7) one can get that
\begin{equation}
\theta^{\rm max}_{J/\psi}=3.932^{\circ},\,\,\,\,\theta^{\rm max}_{J/\psi}=5.377^{\circ}~{\rm and}\,\,\,\,
\theta^{\rm max}_{J/\psi}=5.507^{\circ}
\end{equation}
at initial resonant antikaon beam momenta of 9.417, 9.965 and 10.026 GeV/c, respectively.
Energy-momentum conservation in the reaction (5), proceeding on a free target proton at rest, tells us
that the kinematically allowed $J/\psi$ meson laboratory momentum $p_{J/\psi}$ and total energy $E_{J/\psi}$
($E_{J/\psi}=\sqrt{m^2_{J/\psi}+p^{2}_{J/\psi}})$
in this reaction vary within the following momentum and energy ranges at given initial antikaon momentum:
\begin{equation}
p^{(2)}_{J/\psi}(0^{\circ}) \le p_{J/\psi} \le p^{(1)}_{J/\psi}(0^{\circ}),
\end{equation}
\begin{equation}
E^{(2)}_{J/\psi}(0^{\circ}) \le E_{J/\psi} \le E^{(1)}_{J/\psi}(0^{\circ}),
\end{equation}
where the quantities $p^{(1,2)}_{J/\psi}(0^{\circ})$ and $E^{(1,2)}_{J/\psi}(0^{\circ})$ are the $J/\psi$
momenta and energies at zero polar charmonium production angle in the laboratory frame.
They are defined as follows [70]:
\begin{equation}
p^{(1,2)}_{J/\psi}(0^{\circ})=
{\gamma_{\rm cm}}E^{*}_{J/\psi}(v_{\rm cm}\pm v^{*}_{J/\psi}),
\end{equation}
\begin{equation}
E^{(1,2)}_{J/\psi}(0^{\circ})=
{\gamma_{\rm cm}}(E^{*}_{J/\psi}\pm v_{\rm cm}p^{*}_{J/\psi}).
\end{equation}
Here,
\begin{equation}
{\gamma_{\rm cm}}=(E_{K^-}+m_p)/\sqrt{s},\,\,\,\, v_{\rm cm}=p_{K^-}/(E_{K^-}+m_p),\,\,\,\,
v^{*}_{J/\psi}=p^{*}_{J/\psi}/E^{*}_{J/\psi},\,\,\,\,
E^{*}_{J/\psi}=\sqrt{m^2_{J/\psi}+p^{*2}_{J/\psi}}.
\end{equation}
The sign "+" in Eqs. (14), (15) corresponds to the first quantities $p^{(1)}_{J/\psi}$, $E^{(1)}_{J/\psi}$
and sign "-" - to the second ones $p^{(2)}_{J/\psi}$, $E^{(2)}_{J/\psi}$.
In line with Eqs. (14), (15), for $p_{K^-}=9.417$ GeV/c
the inequalities (12), (13) are:
\begin{equation}
5.695~\le p_{J/\psi}~\le 7.898~{\rm GeV/c},
\end{equation}
\begin{equation}
6.483~\le E_{J/\psi}~\le 8.484~{\rm GeV}.
\end{equation}
For $p_{K^-}=9.965$ GeV/c we obtain:
\begin{equation}
5.488~\le p_{J/\psi}~\le 8.668~{\rm GeV/c},
\end{equation}
\begin{equation}
6.300~\le E_{J/\psi}~\le 9.205~{\rm GeV}.
\end{equation}
And for $p_{K^-}=10.026$ GeV/c we also have:
\begin{equation}
5.471~\le p_{J/\psi}~\le 8.748~{\rm GeV/c},
\end{equation}
\begin{equation}
6.286~\le E_{J/\psi}~\le 9.280~{\rm GeV}.
\end{equation}
Obviously, the binding of target protons and their Fermi motion will distort the distributions of the
outgoing high-momentum (and high-energy) $J/\psi$ mesons
\footnote{$^)$And $\Lambda$ hyperons, for which the kinematically allowed momentum interval looks like
$1.286~\le p_{\Lambda}~\le 4.522~{\rm GeV/c}$, for instance, for $p_{K^-}=10$ GeV/c [70].}$^)$
as well as lead to a wider accessible momentum and energy intervals compared to those given
above by Eqs. (17)--(22). With this, we will ignore the modification of the final high-momentum
$J/\psi$ meson and $\Lambda$ hyperon in the nuclear medium in the present work in the case when
the reaction $K^-p \to {J/\psi}\Lambda$ proceeds on a proton embedded in a nuclear target.

For our purposes, we need the $J/\psi$ energy spectrum
$d\sigma_{{K^-}p \to {J/\psi}{\Lambda}}[\sqrt{s},p_{J/\psi}]/ dE_{J/\psi}$, arising
from this reaction, taking place on a free target proton at rest, as a function of
the $J/\psi$ total energy $E_{J/\psi}$ belonging to the interval (13). It was calculated [70] to be:
\begin{equation}
\frac{d\sigma_{{K^-}p \to {J/\psi}{\Lambda}}[\sqrt{s},p_{J/\psi}]}{dE_{J/\psi}}=
\left(\frac{2{\pi}\sqrt{s}}{p_{K^-}p^{*}_{J/\psi}}\right)
\frac{d\sigma_{{K^{-}}p \to {J/\psi}{\Lambda}}[\sqrt{s},\theta_{J/\psi}^*(x_0)]}{d{\bf \Omega}_{J/\psi}^*}~{\rm for}
~E^{(2)}_{J/\psi}(0^{\circ}) \le E_{J/\psi} \le E^{(1)}_{J/\psi}(0^{\circ}).
\end{equation}
Here, $d\sigma_{{K^{-}}p \to {J/\psi}{\Lambda}}(\sqrt{s},\theta_{J/\psi}^*)/d{\bf \Omega}_{J/\psi}^*$
is the on-shell differential cross section for the production of $J/\psi$ meson in reaction (5) under the
polar angle $\theta_{J/\psi}^*$ in the $K^-p$ c.m.s. This cross section is assumed to have the standard
exponential $t$-shape form, multiplied by the free total cross section
$\sigma_{{K^-}p \to {J/\psi}\Lambda}(\sqrt{s})$ of the reaction ${K^-}p \to {J/\psi}\Lambda$.
The latter one assumes the form [70]:
\begin{equation}
 \sigma_{{K^-}p \to {J/\psi}\Lambda}(\sqrt{s})=7\cdot10^{-6}\cdot
 \sigma_{{K^-}p \to {\phi}\Lambda}(\sqrt{{\tilde s}}),
\end{equation}
where the quantity $\sigma_{{K^-}p \to {\phi}\Lambda}(\sqrt{{\tilde s}})$ represents the free total cross
section of the process ${K^-}p \to {\phi}\Lambda$ and
the center-of-mass collision energies $\sqrt{s}$ and $\sqrt{{\tilde s}}$ are linked by the relation:
\begin{equation}
\sqrt{{\tilde s}}=\sqrt{s}-m_{J/\psi}+m_{\phi}.
\end{equation}
Here, the quantity $m_{\phi}$ is the vacuum $\phi$ meson mass.
For the cross section $\sigma_{{K^-}p \to {\phi}\Lambda}(\sqrt{{\tilde s}})$
we have adopted the following parametrization:
\begin{equation}
\sigma_{{K^-}p \to {\phi}\Lambda}(\sqrt{{\tilde s}})=Bp^*_{\phi}{\rm e}^{-{\beta}p^*_{\phi}},
\end{equation}
suggested in Ref. [74].
Here, the $\phi$ c.m. momentum $p^*_{\phi}$ is defined by the expression:
\begin{equation}
p_{\phi}^*=\frac{1}{2\sqrt{{\tilde s}}}\lambda({\tilde s},m_{\phi}^{2},m_{\Lambda}^2)
\end{equation}
and parameters $B$ and $\beta$ are: $B=315.31$ $\mu$b/(GeV/c), $\beta=1.45$ (GeV/c)$^{-1}$.
The expressions (23)--(27) will be used by us for calculating the free space $J/\psi$ energy spectrum
for incident resonant antikaon beam momenta of 9.417, 9.965 and 10.026 GeV/c (see below) and the $J/\psi$
excitation function near threshold.

As was noticed above (see Eq. (11)), the $J/\psi$ mesons are produced at small angles with respect to
the $K^-$ beam direction at the considered initial $K^-$ meson momenta.
Therefore, in what follows, for calculation of the momentum differential cross section for $J/\psi$ meson production
in ${K^-}A$ interactions from the direct process (5) we will use Eq. (33) from Ref. [70], which now reads:
\begin{equation}
\frac{d\sigma_{{K^-}A\to {J/\psi}X}^{({\rm dir})}
(p_{K^-},p_{J/\psi})}{dp_{J/\psi}}=
2{\pi}\left(\frac{Z}{A}\right)I_{V}[A,\sigma_{{J/\psi}N}]
\int\limits_{\cos20^{\circ}}^{1}d\cos{{\theta_{J/\psi}}}
\left<\frac{d\sigma_{{K^-}p\to {J/\psi}{\Lambda}}(p_{K^-},
p_{J/\psi},\theta_{J/\psi})}{dp_{J/\psi}d{\bf \Omega}_{J/\psi}}\right>_A,
\end{equation}
where
\begin{equation}
I_{V}[A,\sigma]=2{\pi}A\int\limits_{0}^{R}r_{\bot}dr_{\bot}
\int\limits_{-\sqrt{R^2-r_{\bot}^2}}^{\sqrt{R^2-r_{\bot}^2}}dz
\rho(\sqrt{r_{\bot}^2+z^2})
\end{equation}
$$
\times
\exp{\left[-A\sigma_{K^-N}^{\rm tot}\int\limits_{-\sqrt{R^2-r_{\bot}^2}}^{z}
\rho(\sqrt{r_{\bot}^2+x^2})dx
-A{\sigma}\int\limits_{z}^{\sqrt{R^2-r_{\bot}^2}}
\rho(\sqrt{r_{\bot}^2+x^2})dx\right]}.
$$
Here in Eq. (28),
$\left<\frac{d\sigma_{{K^-}p\to {J/\psi}\Lambda}(p_{K^-},
p_{J/\psi},\theta_{J/\psi})}{dp_{J/\psi}d{\bf \Omega}_{J/\psi}}\right>_A$ is the
off-shell inclusive differential cross section for the production of $J/\psi$ mesons
with momentum ${\bf p}_{J/\psi}$ in reaction (5), averaged over the Fermi motion and binding energy
of the protons in the nucleus (cf. Eq. (5) from Ref. [70]).
The quantities $\rho(r)$ as well as $\sigma_{{K^-}N}^{\rm tot}$ and $\sigma=\sigma_{{J/\psi}N}$,
entering into Eq. (29), denote, respectively, the target nucleon density, normalized to unity, as well as
the total cross section of the free ${K^-}N$ interaction and ${J/\psi}N$ absorption cross section
\footnote{$^)$For which we use the values $\sigma_{{K^-}N}^{\rm tot}=22$ mb and $\sigma_{{J/\psi}N}=3.5$ mb [70].}$^)$
.
In our calculations we assume that the neutron and proton
densities have the same radial shape $\rho(r)$. For it we have assumed an harmonic oscillator and
a two-parameter Fermi distributions for $^{12}$C and $^{184}$W target nuclei, respectively.
For the rest of notation see Ref. [70].

\subsection*{2.2. Two-step resonant $J/\psi$ production processes}

At $K^-$ meson beam momenta below 10.2 GeV/c of interest, incident antikaons can produce the neutral
$P_{cs}(4338)^0$, $P_{cs}(4455)^0$ and $P_{cs}(4468)^0$ pentaquark resonances with single strangeness $S=-1$
directly in the first inelastic collisions with intranuclear protons:
\begin{eqnarray}
{K^-}+p \to P_{cs}(4338)^0,\nonumber\\
{K^-}+p \to P_{cs}(4455)^0,\nonumber\\
{K^-}+p \to P_{cs}(4468)^0.
\end{eqnarray}
Then the produced pentaquark resonances can decay into the final state ${J/\psi}\Lambda$,
which will additionally contribute to the $J/\psi$ yield in the ($K^-$,$J/\psi$) reactions
on protons and nuclei:
\begin{eqnarray}
P_{cs}(4338)^0 \to J/\psi+\Lambda,\nonumber\\
P_{cs}(4455)^0 \to J/\psi+\Lambda,\nonumber\\
P_{cs}(4468)^0 \to J/\psi+\Lambda.
\end{eqnarray}
As above in the case of direct $J/\psi$ meson production, at first, we consider here the charmonium
production in the production/decay sequences (30)/(31), taking place on a free target proton at rest.
According to [70], the free space total cross sections
$\sigma_{{K^-}p \to P_{csi}^0 \to {J/\psi}\Lambda}(\sqrt{s},\Gamma_{csi})$
for resonant charmonium production in these sequences can be represented as follows
\footnote{$^)$Here, $i=$1, 2, 3 and $P^0_{cs1}$, $P^0_{cs2}$ and $P^0_{cs3}$
stand for $P_{cs}(4338)^0$, $P_{cs}(4455)^0$ and $P_{cs}(4468)^0$ states, respectively.
Analogously, $\Gamma_{cs1}$, $\Gamma_{cs2}$ and $\Gamma_{cs3}$ will denote in this exploratory study,
correspondingly, the central values of their total decay widths, i.e.: $\Gamma_{cs1}=7.0$ MeV, $\Gamma_{cs2}=7.5$ MeV
and $\Gamma_{cs3}=5.2$ MeV.}$^)$
:
\begin{equation}
\sigma_{{K^-}p \to P_{csi}^0 \to {J/\psi}\Lambda}(\sqrt{s},\Gamma_{csi})=
\sigma_{{K^-}p \to P_{csi}^0}(\sqrt{s},\Gamma_{csi})\theta[\sqrt{s}-(m_{J/\psi}+m_{\Lambda})]
Br[P_{csi}^0 \to {J/\psi}\Lambda].
\end{equation}
Here, $\theta(x)$ is the step function and the quantities $Br[P_{csi}^0 \to {J/\psi}\Lambda]$ ($i=$1, 2, 3)
represent the branching ratios of the decays (31).
In Eq. (32), the quantities $\sigma_{{K^-}p \to P_{csi}^0}(\sqrt{s},\Gamma_{csi})$ are the total cross sections for
production of the $P_{csi}^0$ resonances with the possible spin-parity quantum numbers $J^P=(1/2)^-$ for $P_{cs1}^0$ and $P_{cs2}^0$, and $J^P=(3/2)^-$ for $P_{cs3}^0$ in reactions (30)
\footnote{$^)$Which might be assigned to them within
the hadronic molecular scenario for their internal structure (see Introduction section above).}$^)$
.
These cross sections can be described, using the spectral functions $S_{csi}^0(\sqrt{s},\Gamma_{csi})$
of resonances and knowing the branching fractions $Br[P^0_{csi} \to {K^-}p]$ of their decays to the $K^-p$ mode,
as follows [70]:
\begin{equation}
\sigma_{{K^-}p \to P_{csi}^0}(\sqrt{s},\Gamma_{csi})=
f_{csi}\left(\frac{\pi}{p^*_{K^-}}\right)^2
Br[P_{csi}^0 \to {K^-}p]S_{csi}^0(\sqrt{s},\Gamma_{csi})\Gamma_{csi}, \,\,i=1,2,3,
\end{equation}
where the c.m. 3-momentum in the incoming ${K^-}p$ channel, $p^*_{K^-}$,
is defined by the formula
\begin{equation}
p_{K^-}^*=\frac{1}{2\sqrt{{s}}}\lambda({s},m_{K^-}^{2},m_{p}^2)
\end{equation}
and the ratios of the spin factors $f_{cs1}=2$, $f_{cs2}=2$, $f_{cs3}=4$.
In line with Ref. [70], we suppose that the free spectral functions
$S_{csi}^0(\sqrt{s},\Gamma_{csi})$ of the intermediate $P_{csi}^{0}$ resonances, entering into Eq. (33),
are described by the non-relativistic Breit-Wigner distributions
\footnote{$^)$In view of the fact that the $P_{cs}(4338)^0$ state is very close to the
$\Xi_c^0{\bar D}^{0}$ and $\Xi_c^+D^{-}$ thresholds, its lineshape may be distorted from the conventional
BW distribution due to the double threshold effects [29, 75]. But since the possible pole positions for the
$P_{cs}(4338)^0$ (what concerns the masses and widths), found in [29, 75], are close (within the uncertainties)
to those BW parameters determined in the LHCb analysis, the generation of the $P_{cs}(4338)^0$ shape with BW
one carried out in the present work is reasonable enough for our exploratory purposes.}$^)$
:
\begin{equation}
S_{csi}^0(\sqrt{s},\Gamma_{csi})=
\frac{1}{2\pi}\frac{\Gamma_{csi}}{(\sqrt{s}-M_{csi})^2+{\Gamma}_{csi}^{2}/4}.
\end{equation}
Inspection of Eqs. (32), (33) tells us that to evaluate the $J/\psi$ production cross sections
from the production/decay sequences (30)/(31), taking place on a vacuum proton
(and on a bound in a nuclear medium proton), one needs to know the branching ratios
$Br[P^0_{csi} \to {K^-}p]$ and $Br[P_{csi}^0 \to {J/\psi}\Lambda]$ ($i=$1, 2, 3)
of the $P^0_{csi}$ decays to the $K^-p$ and ${J/\psi}\Lambda$ channels. We focus now on them.

The branching ratios $Br[P_{csi}^0 \to {J/\psi}\Lambda]$ of the decays (31) were not determined
experimentally. Therefore, we must rely on the theoretical predictions for them as well as on the similarity
of the behaviors and decay properties of the $P_{cs}$ and $P_c$ systems (cf. Introduction and Refs. [15, 35]).
Thus, the decay rates of the modes $P_c^+ \to {J/\psi}p$ and $P_{cs}^0 \to {J/\psi}\Lambda^0$ are expressed
in Ref. [15] in terms of the single model parameter $\Lambda$, which should be constrained from the future experiments,
and these rates are comparable to each other.
Model-dependent upper limits on branching fractions $Br[P_{c}(4312)^+ \to {J/\psi}p]$,
$Br[P_{c}(4440)^+ \to {J/\psi}p]$ and $Br[P_{c}(4457)^+ \to {J/\psi}p]$ of several percent
were set by the GlueX experiment [76] at JLab. Preliminary results from a factor of 10 more data,
collected in the $J/\psi$--007 experiment [77] also at JLab, focused on the large $t$ region
\footnote{$^)$In which the rather flat resonant production of $J/\psi$ through the $P_c^+$ in photon-induced
reactions on the protons is expected to be enhanced relative to the suppressed here mostly forward diffractive production.}$^)$
in searching for the LHCb hidden-charm pentaquarks [24], also observe  no signals for them
and set more stringent upper limits on the
cross sections for production of the $P_{c}(4312)^+$, $P_{c}(4440)^+$ and $P_{c}(4457)^+$ states
in ${\gamma}p$ collisions almost an order of magnitude below the respective GlueX limits [76].
With this, with upper limits on the branching fractions $Br[P_{c}(4312/4440/4457)^+ \to {J/\psi}p]$
already available from the GlueX experiment [76] and within the representation of these cross sections,
in which they are proportional, respectively, to the $Br^2[P_{c}(4312/4440/4457)^+ \to {J/\psi}p]$,
the upper limits on the above branching fractions, which are
expected from the $J/\psi$-007 experiment [77], were estimated to be at the level of 1\% in Ref. [78].
Based on the branching ratios and fractions measured by the LHCb and GlueX Collaborations,
the authors of Ref. [79] obtain that a lower
limit of $Br[P^+_{c} \to {J/\psi}p]$ is of the order of 0.05\% $\sim$ 0.5\%.
Taking into account these findings as well as the fact that the ratios of the branching fractions
$Br[P_{c}^+ \to {J/\psi}p]$ between three states $P_c(4312)^+$, $P_c(4440)^+$ and $P_c(4457)^+$
were predicted to be $\sim$ 1 in Ref. [16], we adopted in our study [78] for the branching fractions of
$P^+_{c}\to {J/\psi}p$ decays for each individual $P_c^+$ state three following conservative options:
$Br[P^+_{c} \to {J/\psi}p]=0.25$, 0.5 and 1\%.
With this and in view of the aforesaid, it is natural to assume for all branching ratios
$Br[P_{csi}^0 \to {J/\psi}\Lambda]$ of the decays (31) the same three main options, namely:
$Br[P_{csi}^0 \to {J/\psi}\Lambda]=0.25$, 0.5 and 1\% ($i=$1, 2, 3) as those given above and used for the
$P^+_{c}\to {J/\psi}p$ decays, and additional one with reduced value of 0.125\% of these ratios
in order to see better the size of their impact on the resonant $J/\psi$ yield in ${K^-}p \to {J/\psi}\Lambda$,
${K^-}$$^{12}$C $\to {J/\psi}X$ and ${K^-}$$^{184}$W $\to {J/\psi}X$ reactions
\footnote{$^)$It should be pointed out that in view of the aforementioned these options are more realistic than
those of 3, 5, 10, 15 and 50\% employed in Ref. [70] for the $P_{cs}(4459)^0 \to {J/\psi}\Lambda$ decays.}$^)$
.
It is worth mentioning that the investigations in Refs. [34, 54] support in some sense such choice for the
branching fractions of the decays (31).
Thus, investigation of the strong decays of the $P_{cs}(4459)^0$ pentaquark, assuming that it is a pure $S$-wave
$\Xi_c{\bar D}^*$ molecular state with two possible spin-parity assignments $J^P=(1/2)^-$ and $J^P=(3/2)^-$,
has been performed in Ref. [34] through hadronic loops with the help of the effective Lagrangians.
In comparison with the LHCb data [24], the $S$-wave $\Xi_c{\bar D}^*$ molecule with $J^P=(1/2)^-$ assignment
for the $P_{cs}(4459)^0$ is supported by the study [34]. Using the results for the total decay width and for
the partial width of the decay of the $P_{cs}(4459)^0$ with $J^P=(1/2)^-$ into the ${J/\psi}\Lambda$,
the branching ratio of this decay could be evaluated at the level of 1\%.
Further, the author in Ref. [54], interpreting the pentaquark $P_{cs}(4459)^0$
observed by the LHCb Collaboration [24] as the hadronic molecule with the dominant
$S$-wave channels $\Xi_c{\bar D}^*$ and $\Xi_c^*{\bar D}$ with the spin-parity combination
$J^P=(3/2)^-$, predicted adopting the one-boson-exchange model that
the partial decay width for the process $P_{cs}(4459)^0 \to {J/\psi}\Lambda$ is around 0.1 MeV in the
resonance region. With this value and with the $P_{cs}(4459)^0$ resonance total decay width of about
20 MeV also predicted in [54] in this region, we obtain that the branching fraction
$Br[P_{cs}(4459)^0 \to {J/\psi}\Lambda]$ amounts to 0.5\%. Moreover, 1\% branching ratio of the
$P_{cs}(4459)^0 \to {J/\psi}\Lambda$ decay was chosen as the main one in Ref. [69]. In previous works [69, 70],
devoted to the study of the $P_{cs}(4459)^0$ resonance production in antikaon-induced reactions on protons and
nuclei, the value of 0.01\% was used for the branching fraction $Br[P_{cs}(4459)^0 \to {K^-}p]$. It was chosen
to be similar to that characterizing [80] the $P_c^0 \to {\pi^-}p$ decay. Based on the molecular scenario, the
partial decay widths of the $P_{cs}(4459)^0$ with $J^P=(1/2)^-$ and $J^P=(3/2)^-$ into $K^-p$ final state via hadronic
loops were evaluated in Ref. [71] to be of the order of 2.05 and 0.24 KeV, respectively. If we adopt an experimental
magnitude of 17.3 MeV for the $P_{cs}(4459)^0$ total decay width, we obtain with the above partial decay widths that
the branching fraction $Br[P_{cs}(4459)^0 \to {K^-}p]$ would be $\sim$ 0.01\% and 0.001\% for
the $P_{cs}(4459)^0$ with $J^P=(1/2)^-$ and $J^P=(3/2)^-$, correspondingly. In line with the aforementioned, it is
also natural to employ in our cross-section calculations for the branching ratios $Br[P_{csi}^0 \to {K^-}p]$
($i=$1, 2, 3) the main value of 0.01\%. To see the effect of the branching ratios on the resonant $J/\psi$ yield
in the reactions of interest, we will also use for them in these calculations the value of 0.001\%.

Before going further, we determine the resonant $J/\psi$ energy distributions\\
$d\sigma_{{K^-}p \to P_{csi}^0 \to {J/\psi}\Lambda}[\sqrt{s},p_{J/\psi}]/ dE_{J/\psi}$ ($i=$1, 2, 3)
from the two-step processes (30)/(31), proceeding on the free target proton at rest,
in addition to that (23) from the background ${K^-}p \to {J/\psi}\Lambda$ reaction.
The energy-momentum conservation in these precesses
leads to the conclusion that the kinematical characteristics
of $J/\psi$ mesons produced in them and in this reaction are the same at given incident $K^-$
meson momentum. The results of the calculations, performed in Ref. [70]
assuming that the decays of the $P_{csi}^0$ to the ${J/\psi}\Lambda$ are dominated
by the lowest partial waves with relative orbital angular momentum $L=0$,
show that these distributions
can be represented in the following way:
\begin{equation}
\frac{d\sigma_{{K^-}p \to P_{csi}^0 \to {J/\psi}\Lambda}[\sqrt{s},p_{J/\psi}]}{dE_{J/\psi}}=
\sigma_{{K^-}p \to P_{csi}^0}(\sqrt{s},\Gamma_{csi})\theta[\sqrt{s}-(m_{J/\psi}+m_{\Lambda})]\times
\end{equation}
$$
\times
\left(\frac{\sqrt{s}}{2p_{K^-}p^{*}_{J/\psi}}\right)
Br[P_{csi}^0 \to {J/\psi}\Lambda]~{\rm for}
~E^{(2)}_{J/\psi}(0^{\circ}) \le E_{J/\psi} \le E^{(1)}_{J/\psi}(0^{\circ}).
$$
Eq. (36) shows that the free space $J/\psi$ energy distributions, which arise from the
production/decay chains (30)/(31), exhibit a totally flat behavior within the
allowed energy range (13).

For calculation of the $J/\psi$ inclusive differential cross section (momentum spectrum) arising from the
production and decay of the intermediate resonances $P_{csi}^0$ in $K^-A$ collisions we will use
the following expression based on Eq. (78) from Ref. [70]:
\begin{equation}
\frac{d\sigma_{{K^-}A\to {J/\psi}X}^{({\rm sec,i})}(p_{K^-},p_{J/\psi})}{dp_{J/\psi}}=
2{\pi}\left(\frac{Z}{A}\right)I_{V}[A,\sigma^{\rm in}_{{P_{csi}}N}]
\end{equation}
$$
\times
\int\limits_{\cos20^{\circ}}^{1}d\cos{{\theta_{J/\psi}}}
\left<\frac{d\sigma_{{K^-}p \to P_{csi}^0 \to {J/\psi}{\Lambda}}(p_{K^-},
p_{J/\psi},\theta_{J/\psi})}{dp_{J/\psi}d{\bf \Omega}_{J/\psi}}\right>_A,\,\,\,\,i=1, 2, 3;
$$
where
$\left<\frac{d\sigma_{{K^-}p \to P_{csi}^0 \to {J/\psi}{\Lambda}}(p_{K^-},
p_{J/\psi},\theta_{J/\psi})}{dp_{J/\psi}d{\bf \Omega}_{J/\psi}}\right>_A$ is the
off-shell inclusive differential cross section for the production of $J/\psi$ mesons
with momentum ${\bf p}_{J/\psi}$ in production/decay chain ${K^-}p \to P_{csi}^0 \to {J/\psi}{\Lambda}$,
averaged over the Fermi motion and binding energy of the protons in the nucleus (cf. Eq. (71) from Ref. [70]).
The quantity $I_{V}[A,\sigma^{\rm in}_{P_{csi}N}]$ in Eq. (37) is defined above by Eq. (29), in which one
needs to make the substitution $\sigma \to \sigma^{\rm in}_{P_{csi}N}$.
Here, $\sigma^{\rm in}_{P_{csi}N}$ is the $P_{csi}^0$--nucleon inelastic total cross section.
For this cross section we will use in our calculations the same value of 23.7 mb as that adopted in Ref. [70]
for the inelastic cross section $\sigma^{\rm in}_{P_{cs}(4459)^0N}$ and obtained here in the molecular picture of
the $P_{cs}(4459)^0$.
\begin{figure}[h!]
\begin{center}
\includegraphics[width=16.0cm]{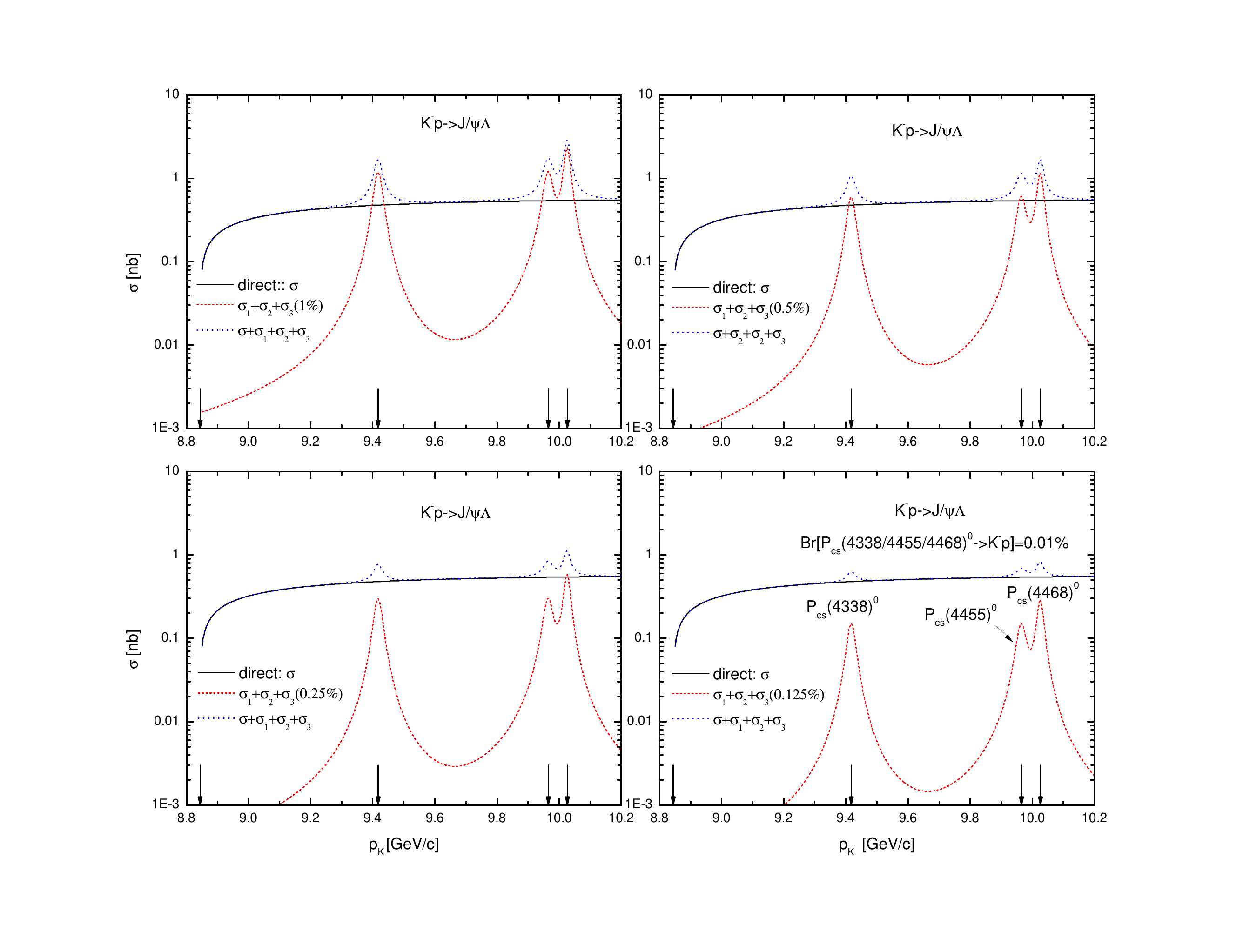}
\vspace*{-2mm} \caption{(Color online.) The non-resonant total cross section $\sigma$ for the reaction
${K^-}p \to {J/\psi}\Lambda$ (black solid curve), calculated on the basis of Eqs. (24)--(27).
Incoherent sum (blue dotted curve) of it and the sum
$\sigma_1$+$\sigma_2$+$\sigma_3$ (red short-dashed curve) of the total cross sections $\sigma_1$,
$\sigma_2$ and $\sigma_3$ for the
resonant $J/\psi$ production in the processes ${K^-}p \to P_{cs}(4338)^0 \to {J/\psi}\Lambda$,
${K^-}p \to P_{cs}(4455)^0 \to {J/\psi}\Lambda$
and ${K^-}p \to P_{cs}(4468)^0 \to {J/\psi}\Lambda$, calculated in line with Eq. (32)
assuming that the resonances
$P_{cs}(4338)^0$, $P_{cs}(4455)^0$ and $P_{cs}(4468)^0$ with the spin-parity quantum
numbers $J^P=(1/2)^-$, $J^P=(1/2)^-$ and $J^P=(3/2)^-$
decay to the $K^-p$ and ${J/\psi}\Lambda$ channels with all three individual branching fractions 0.01\%
and 1, 0.5, 0.25 and 0.125\% (respectively, upper left, upper right,
lower left and lower right panels), as functions of the incident $K^-$ meson momentum.
The left and three right arrows indicate, correspondingly, the threshold momentum
$p^{\rm th}_{K^-}=8.844$ GeV/c for the reaction ${K^-}p \to {J/\psi}\Lambda$, taking place on a free target
proton being at rest, and the resonant antikaon momenta $p^{\rm R1}_{K^-}=9.417$ GeV/c, $p^{\rm R2}_{K^-}=9.965$ GeV/c
and $p^{\rm R3}_{K^-}=10.026$ GeV/c.}
\label{void}
\end{center}
\end{figure}
\begin{figure}[h!]
\begin{center}
\includegraphics[width=16.0cm]{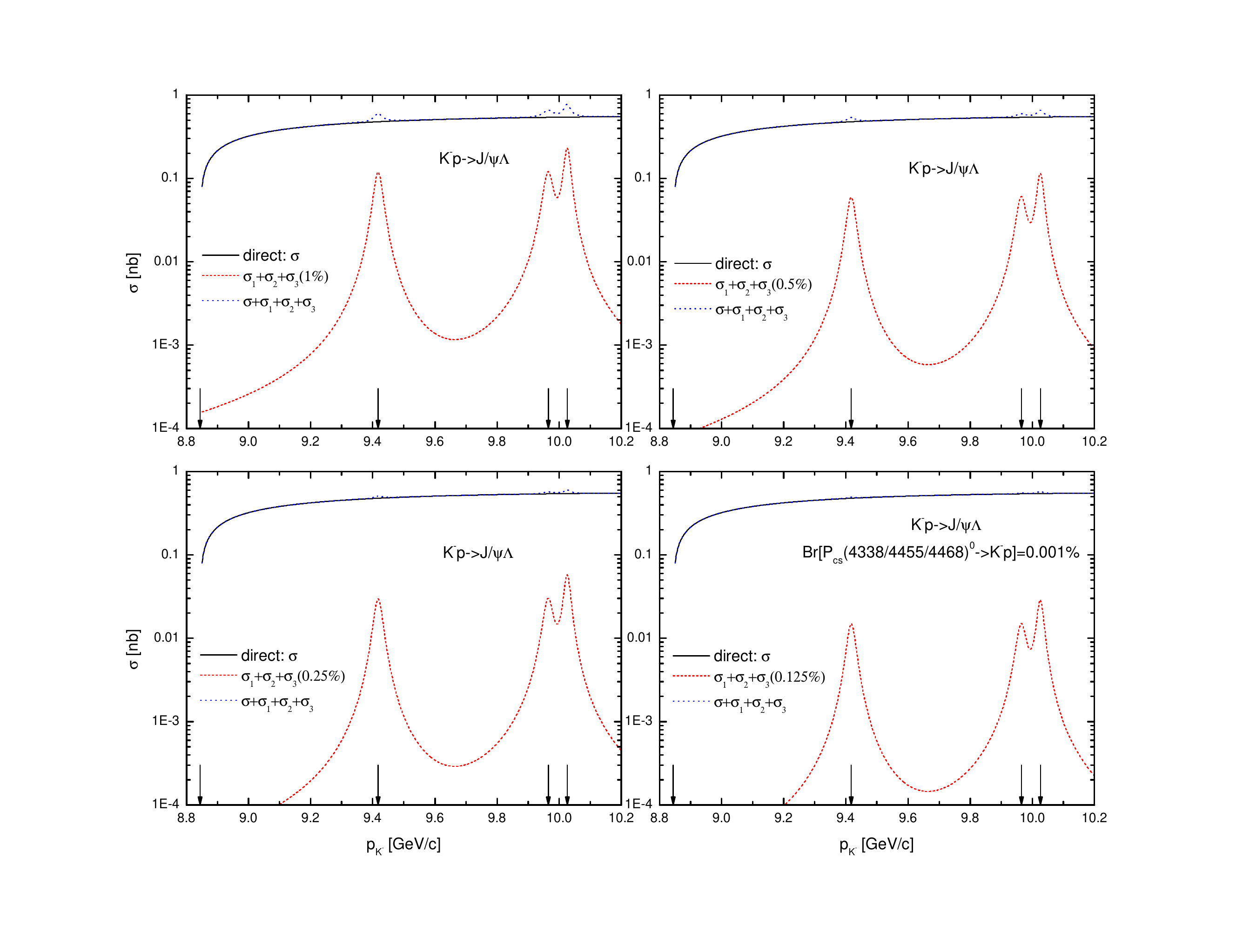}
\vspace*{-2mm} \caption{(Color online.) The same as in Fig. 1, but calculated for all three branching ratios of
$P^0_{csi}$ ($i=1$, 2, 3) decays to the $K^-p$ mode of 0.001\%.}
\label{void}
\end{center}
\end{figure}
\begin{figure}[!h]
\begin{center}
\includegraphics[width=16.0cm]{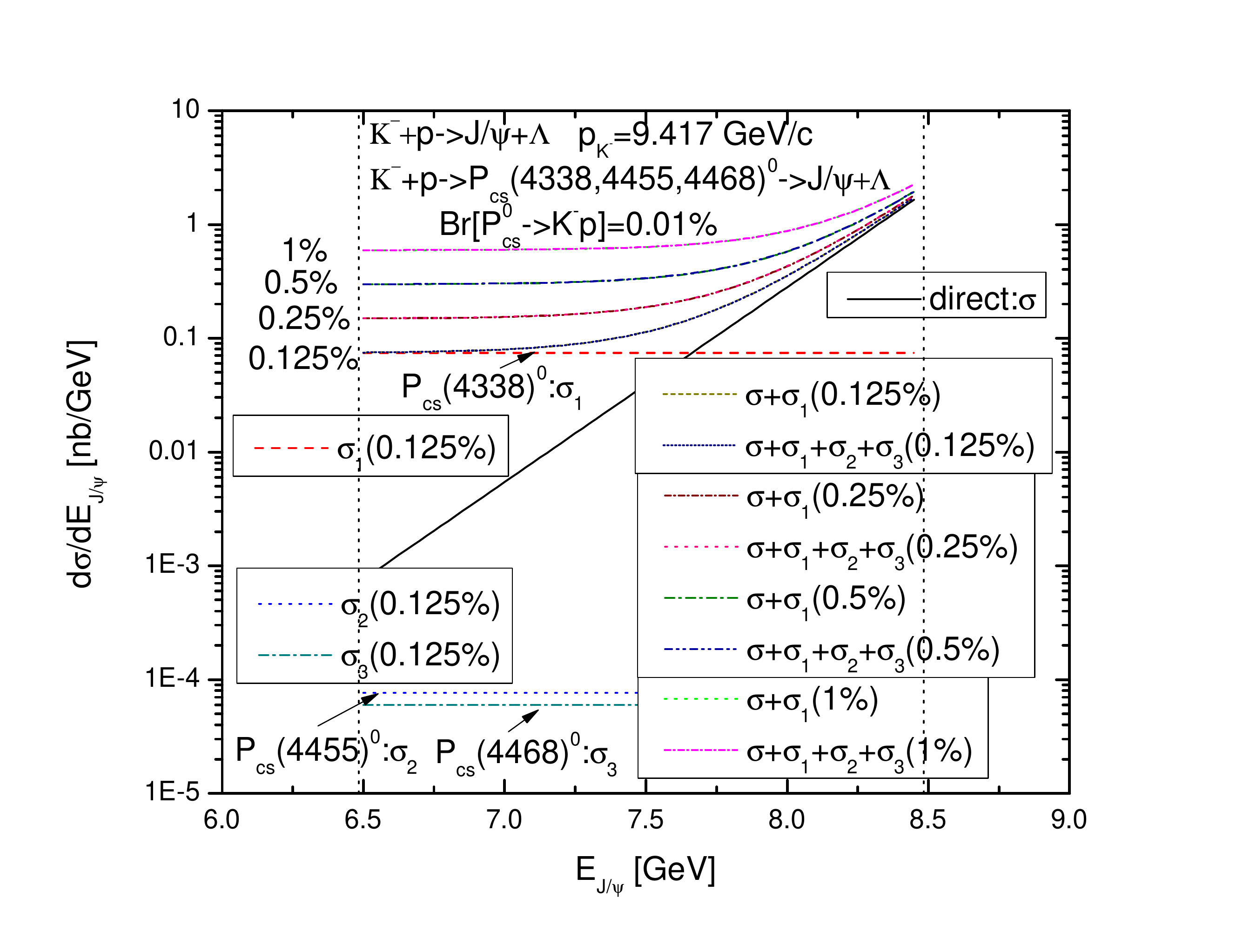}
\vspace*{-2mm} \caption{(Color online.) The direct non-resonant $J/\psi$ energy distribution in the free space
elementary process ${K^-}p \to {J/\psi}\Lambda$,
calculated in line with Eq. (23) at initial $K^-$ meson resonant momentum of 9.417 GeV/c
in the laboratory system (black solid curve). The resonant $J/\psi$ energy distributions in the two-step processes
${K^-}p \to P_{cs}(4338)^0 \to {J/\psi}\Lambda$,
${K^-}p \to P_{cs}(4455)^0 \to {J/\psi}\Lambda$ and ${K^-}p \to P_{cs}(4468)^0 \to {J/\psi}\Lambda$,
calculated in line with Eq. (36) at the same incident antikaon momentum of 9.417 GeV/c
assuming that the resonances $P_{cs}(4338)^0$, $P_{cs}(4455)^0$, $P_{cs}(4468)^0$
with the spin-parity assignments $J^P=(1/2)^-$, $J^P=(1/2)^-$, $J^P=(3/2)^-$, correspondingly,
all decay to the $K^-p$ and ${J/\psi}\Lambda$ modes
with branching fractions of 0.01\% and 0.125\% (respectively, red dashed, blue dotted, dark cyan dashed-doted curves).
Incoherent sum of the direct non-resonant $J/\psi$ energy distribution and resonant ones, calculated supposing
that the resonances $P_{cs}(4338)^0$ as well as $P_{cs}(4338)^0$, $P_{cs}(4455)^0$, $P_{cs}(4468)^0$
with the same spin-parity combinations all decay to the $K^-p$ and ${J/\psi}\Lambda$ with branching
fractions 0.01\% and 0.125, 0.25, 0.5, 1\% (respectively, dark yellow short-dashed,
wine short-dashed-dotted, olive dashed-dotted, green dotted as well as navy short-dotted, pink dotted, royal dashed-dotted-dotted, magenta  short-dashed-dotted curves), all as functions of the total $J/\psi$ energy
$E_{J/\psi}$ in the laboratory system.
The vertical dotted lines indicate the range of $J/\psi$ allowed energies in this system
($6.483 \le E_{J/\psi} \le 8.484~{\rm GeV}$)
for the considered direct non-resonant and resonant $J/\psi$ production off a free target proton at rest at
given initial $K^-$ meson resonant momentum of 9.417 GeV/c.}
\label{void}
\end{center}
\end{figure}
\begin{figure}[!h]
\begin{center}
\includegraphics[width=16.0cm]{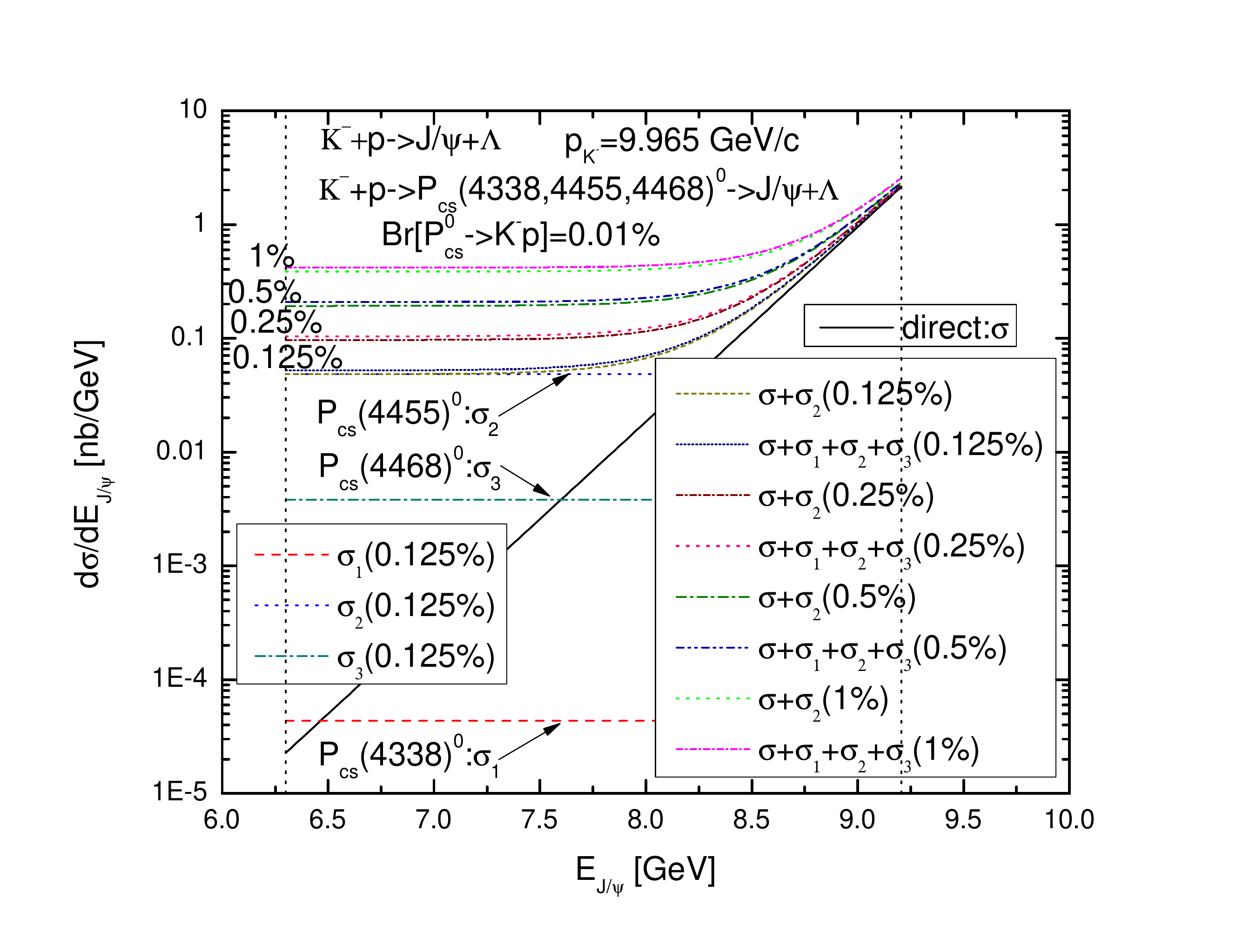}
\vspace*{-2mm} \caption{(Color online.) The direct non-resonant $J/\psi$ energy distribution in the free space
elementary process ${K^-}p \to {J/\psi}\Lambda$,
calculated in line with Eq. (23) at initial $K^-$ meson resonant momentum of 9.965 GeV/c
in the laboratory system (black solid curve). The resonant $J/\psi$ energy distributions in the two-step processes
${K^-}p \to P_{cs}(4338)^0 \to {J/\psi}\Lambda$,
${K^-}p \to P_{cs}(4455)^0 \to {J/\psi}\Lambda$ and ${K^-}p \to P_{cs}(4468)^0 \to {J/\psi}\Lambda$,
calculated in line with Eq. (36) at the same incident antikaon momentum of 9.965 GeV/c
assuming that the resonances $P_{cs}(4338)^0$, $P_{cs}(4455)^0$, $P_{cs}(4468)^0$
with the spin-parity assignments $J^P=(1/2)^-$, $J^P=(1/2)^-$, $J^P=(3/2)^-$, correspondingly,
all decay to the $K^-p$ and ${J/\psi}\Lambda$ modes
with branching fractions of 0.01\% and 0.125\% (respectively, red dashed, blue dotted, dark cyan dashed-doted curves).
Incoherent sum of the direct non-resonant $J/\psi$ energy distribution and resonant ones, calculated supposing
that the resonances $P_{cs}(4455)^0$ as well as $P_{cs}(4338)^0$, $P_{cs}(4455)^0$, $P_{cs}(4468)^0$
with the same spin-parity combinations all decay to the $K^-p$ and ${J/\psi}\Lambda$ with branching
fractions 0.01\% and 0.125, 0.25, 0.5, 1\% (respectively, dark yellow short-dashed,
wine short-dashed-dotted, olive dashed-dotted, green dotted as well as navy short-dotted, pink dotted, royal dashed-dotted-dotted, magenta  short-dashed-dotted curves), all as functions of the total $J/\psi$ energy
$E_{J/\psi}$ in the laboratory system.
The vertical dotted lines indicate the range of $J/\psi$ allowed energies in this system
($6.300 \le E_{J/\psi} \le 9.205~{\rm GeV}$)
for the considered direct non-resonant and resonant $J/\psi$ production off a free target proton at rest at
given initial $K^-$ meson resonant momentum of 9.965 GeV/c.}
\label{void}
\end{center}
\end{figure}
\begin{figure}[!h]
\begin{center}
\includegraphics[width=16.0cm]{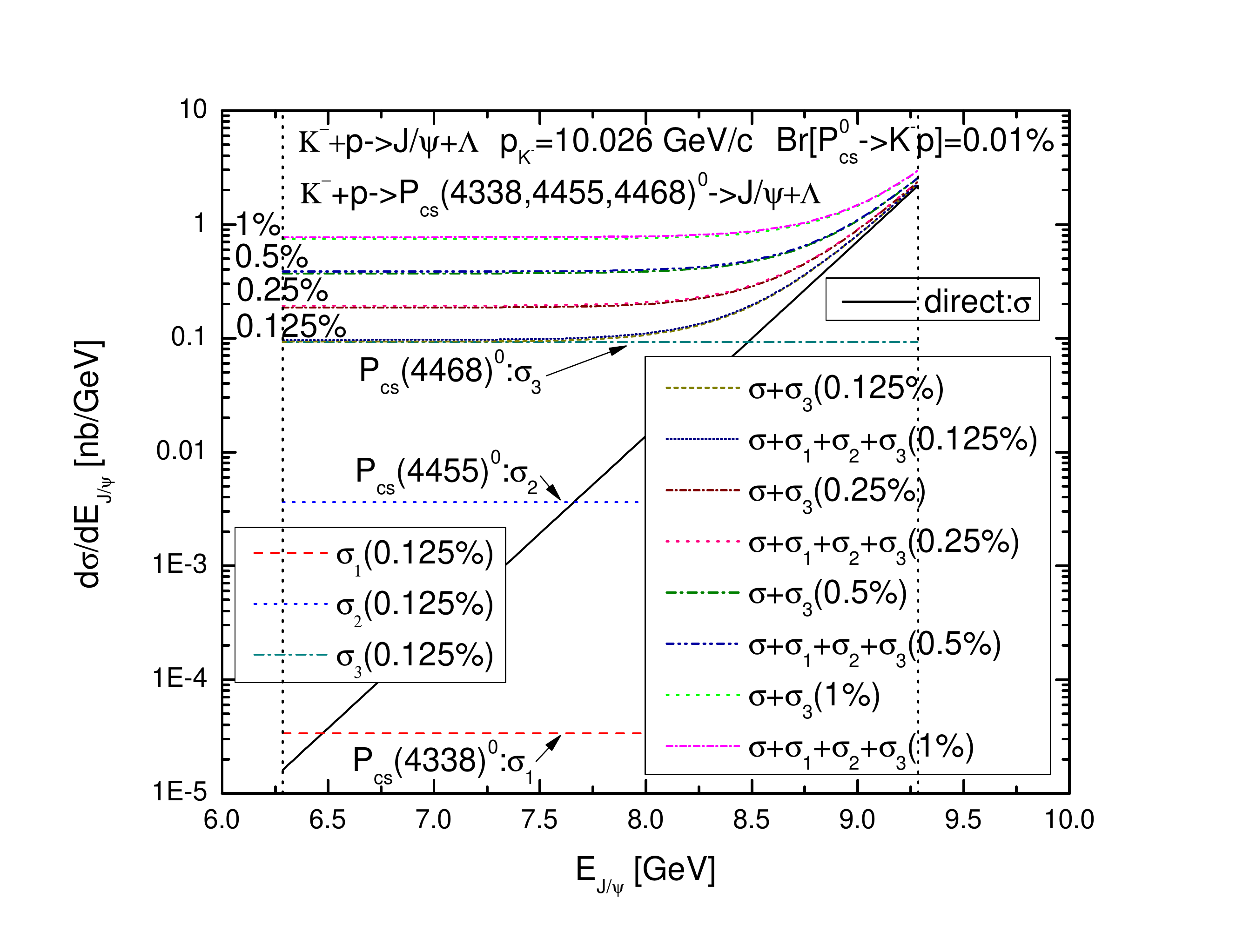}
\vspace*{-2mm} \caption{(Color online.) The direct non-resonant $J/\psi$ energy distribution in the free space
elementary process ${K^-}p \to {J/\psi}\Lambda$,
calculated in line with Eq. (23) at initial $K^-$ meson resonant momentum of 10.026 GeV/c
in the laboratory system (black solid curve). The resonant $J/\psi$ energy distributions in the two-step processes
${K^-}p \to P_{cs}(4338)^0 \to {J/\psi}\Lambda$,
${K^-}p \to P_{cs}(4455)^0 \to {J/\psi}\Lambda$ and ${K^-}p \to P_{cs}(4468)^0 \to {J/\psi}\Lambda$,
calculated in line with Eq. (36) at the same incident antikaon momentum of 10.026 GeV/c
assuming that the resonances $P_{cs}(4338)^0$, $P_{cs}(4455)^0$, $P_{cs}(4468)^0$
with the spin-parity assignments $J^P=(1/2)^-$, $J^P=(1/2)^-$, $J^P=(3/2)^-$, correspondingly,
all decay to the $K^-p$ and ${J/\psi}\Lambda$ modes
with branching fractions of 0.01\% and 0.125\% (respectively, red dashed, blue dotted, dark cyan dashed-doted curves).
Incoherent sum of the direct non-resonant $J/\psi$ energy distribution and resonant ones, calculated supposing
that the resonances $P_{cs}(4468)^0$ as well as $P_{cs}(4338)^0$, $P_{cs}(4455)^0$, $P_{cs}(4468)^0$
with the same spin-parity combinations all decay to the $K^-p$ and ${J/\psi}\Lambda$ with branching
fractions 0.01\% and 0.125, 0.25, 0.5, 1\% (respectively, dark yellow short-dashed,
wine short-dashed-dotted, olive dashed-dotted, green dotted as well as navy short-dotted, pink dotted, royal dashed-dotted-dotted, magenta  short-dashed-dotted curves), all as functions of the total $J/\psi$ energy
$E_{J/\psi}$ in the laboratory system.
The vertical dotted lines indicate the range of $J/\psi$ allowed energies in this system
($6.286 \le E_{J/\psi} \le 9.280~{\rm GeV}$)
for the considered direct non-resonant and resonant $J/\psi$ production off a free target proton at rest at
given initial $K^-$ meson resonant momentum of 10.026 GeV/c.}
\label{void}
\end{center}
\end{figure}
\begin{figure}[htb]
\begin{center}
\includegraphics[width=16.0cm]{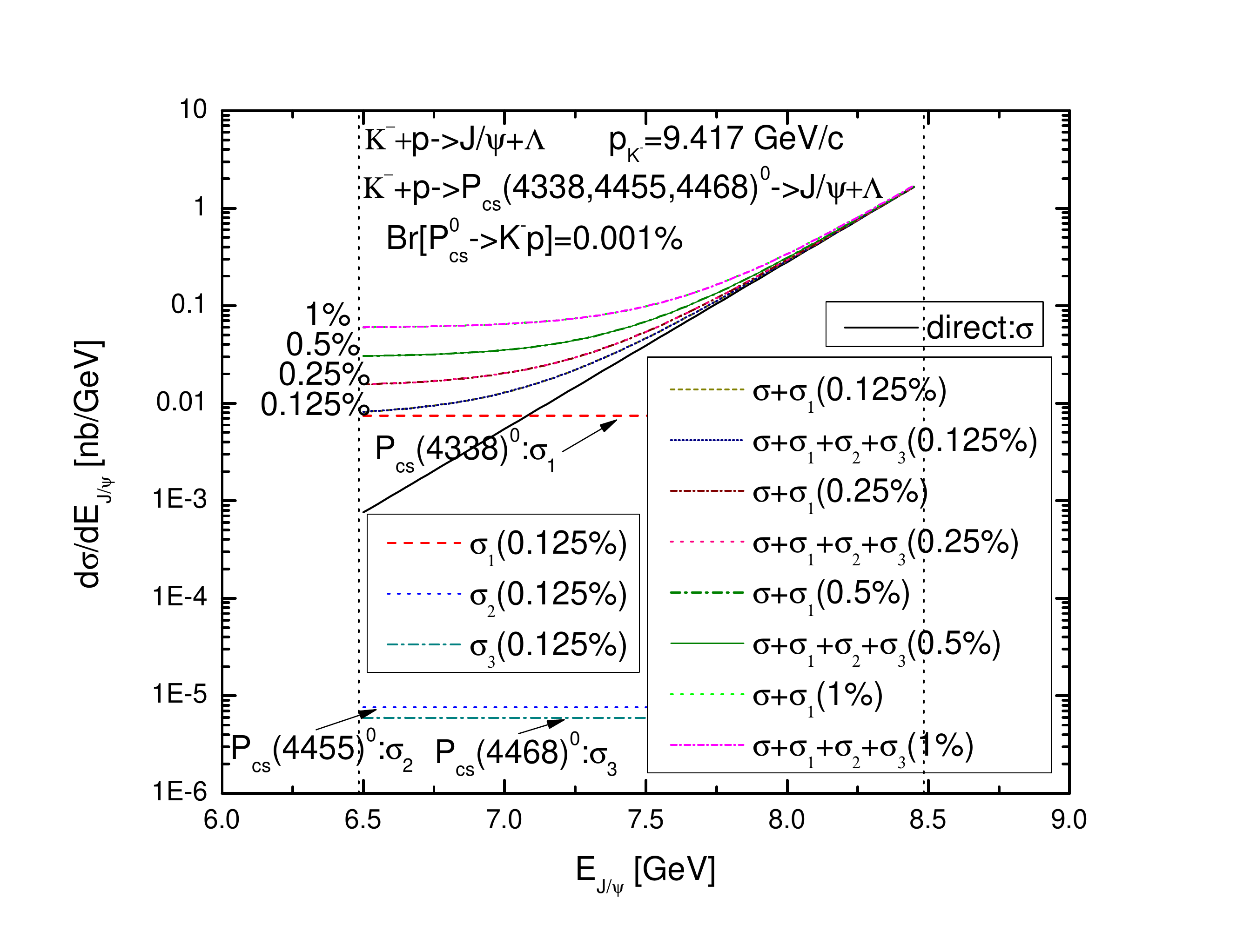}
\vspace*{-2mm} \caption{(Color online.) The same as in Fig. 3, but calculated for all three branching ratios of
$P^0_{csi}$ ($i=1$, 2, 3) decays to the $K^-p$ mode of 0.001\%.}
\label{void}
\end{center}
\end{figure}
\begin{figure}[htb]
\begin{center}
\includegraphics[width=16.0cm]{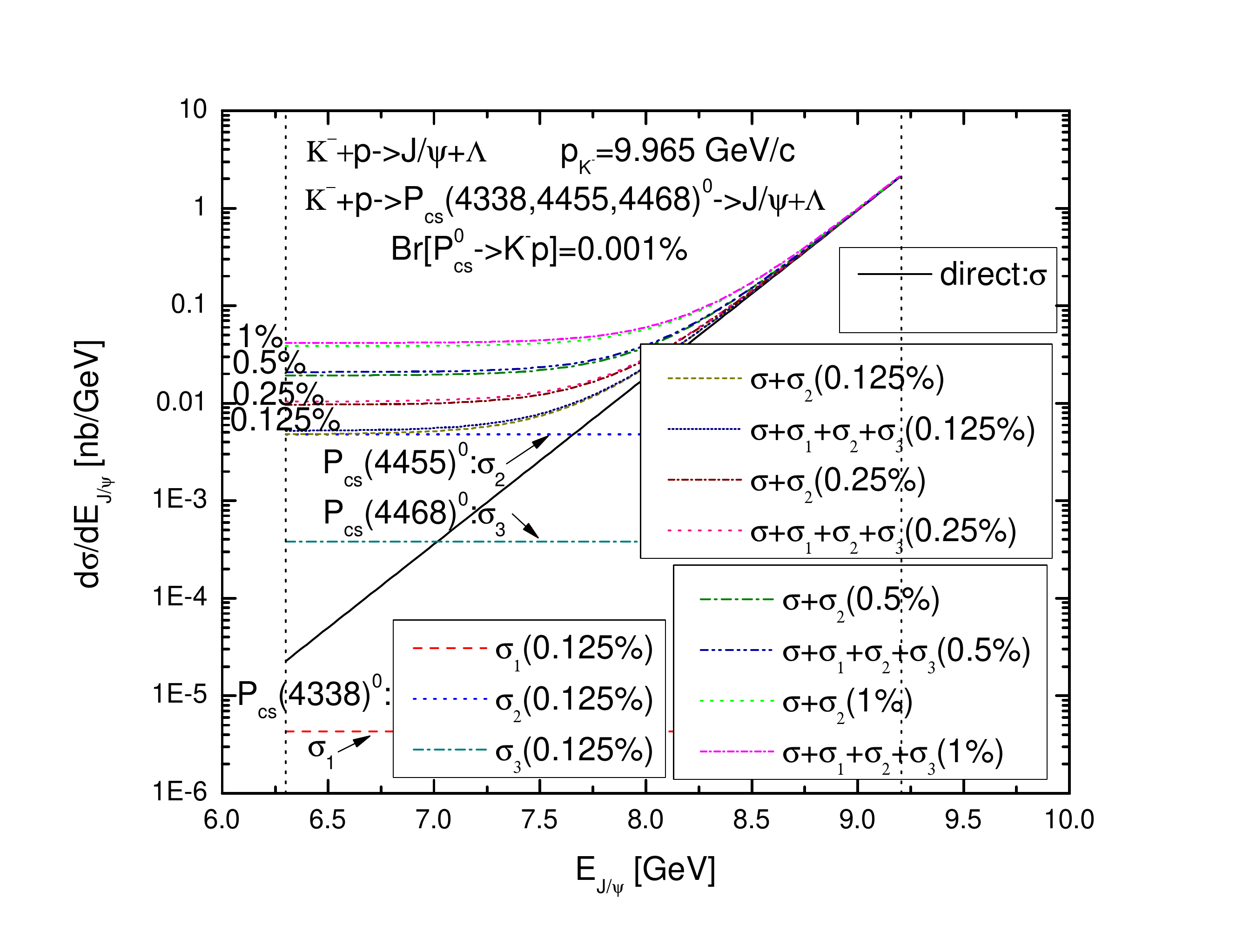}
\vspace*{-2mm} \caption{(Color online.) The same as in Fig. 4, but calculated for all three branching ratios of
$P^0_{csi}$ ($i=1$, 2, 3) decays to the $K^-p$ mode of 0.001\%.}
\label{void}
\end{center}
\end{figure}
\begin{figure}[htb]
\begin{center}
\includegraphics[width=16.0cm]{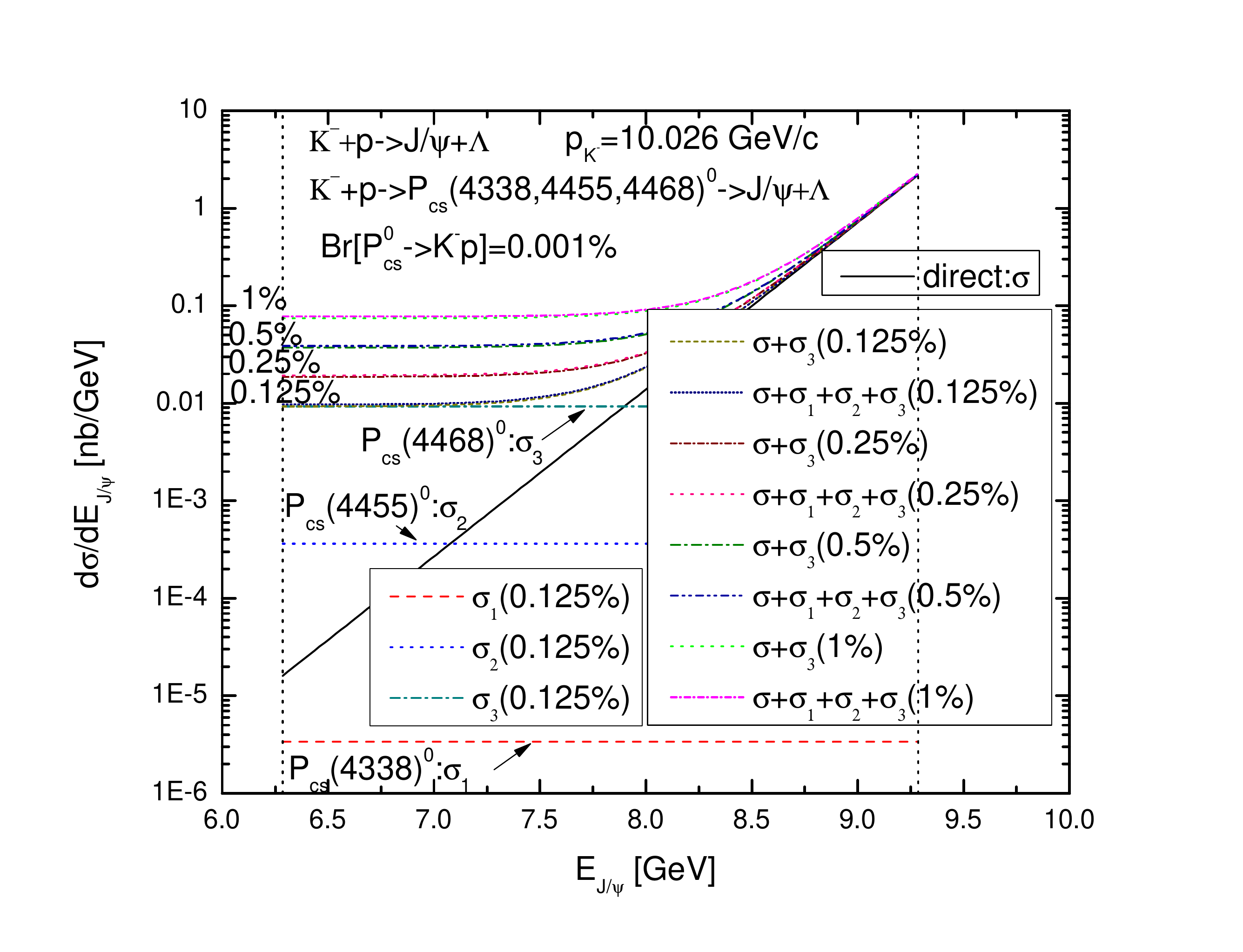}
\vspace*{-2mm} \caption{(Color online.) The same as in Fig. 5, but calculated for all three branching ratios of
$P^0_{csi}$ ($i=1$, 2, 3) decays to the $K^-p$ mode of 0.001\%.}
\label{void}
\end{center}
\end{figure}
\begin{figure}[!h]
\begin{center}
\includegraphics[width=16.0cm]{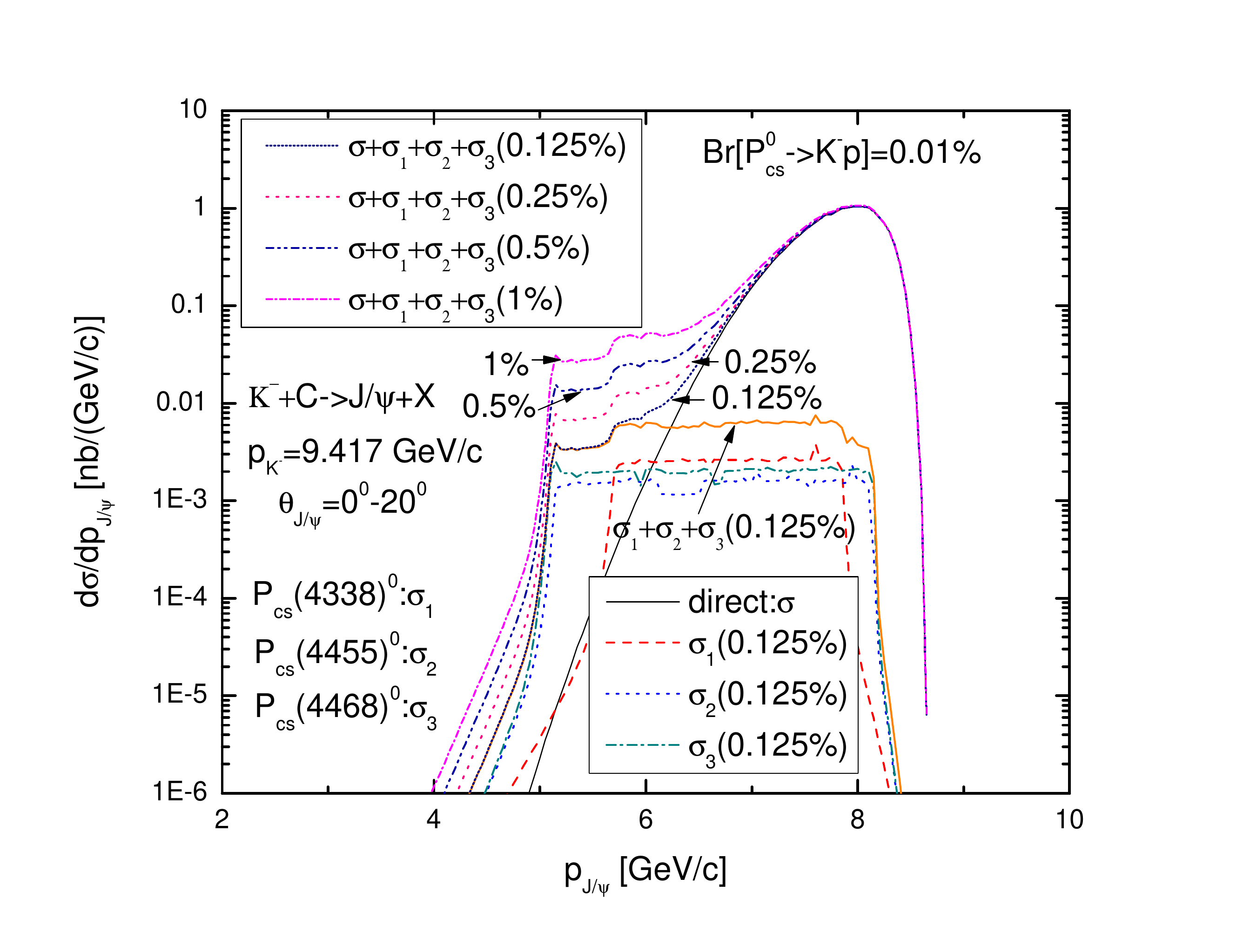}
\vspace*{-2mm} \caption{(Color online.) The direct non-resonant momentum distribution
of $J/\psi$ mesons, produced in the reaction ${K^-}{\rm ^{12}C} \to {J/\psi}X$
in the laboratory polar angular range of 0$^{\circ}$--20$^{\circ}$ and calculated in line with Eq. (28)
at initial antikaon resonant momentum of 9.417 GeV/c in the laboratory system (black solid curve).
The resonant momentum distributions of $J/\psi$ mesons, produced
in the two-step processes ${K^-}p \to P_{cs}(4338)^0 \to {J/\psi}\Lambda$,
${K^-}p \to P_{cs}(4455)^0 \to {J/\psi}\Lambda$ and
${K^-}p \to P_{cs}(4468)^0 \to {J/\psi}\Lambda$ and
calculated in line with Eq. (37) at the same incident antikaon momentum of 9.417 GeV/c
assuming that the resonances $P_{cs}(4338)^0$, $P_{cs}(4455)^0$
and $P_{cs}(4468)^0$ with the spin-parity assignments
$J^P=(1/2)^-$, $J^P=(1/2)^-$ and $J^P=(3/2)^-$, correspondingly,
all decay to the $K^-p$ and ${J/\psi}\Lambda$ modes
with branching fractions 0.01\% and 0.125\% (respectively, red dashed, blue dotted, dark cyan dashed-doted curves)
and their incoherent sum (orange solid curve).
Incoherent sum of the direct non-resonant $J/\psi$ momentum distribution and resonant ones,
calculated supposing that the resonances
$P_{cs}(4338)^0$, $P_{cs}(4455)^0$, $P_{cs}(4468)^0$
with the same spin-parity combinations all decay to the $K^-p$ and ${J/\psi}\Lambda$ with branching
fractions 0.01\% and 0.125, 0.25, 0.5 and 1\% (respectively, navy short-dotted, pink dotted, royal dashed-dotted-dotted and magenta short-dashed-dotted curves), all as functions of the $J/\psi$ momentum $p_{J/\psi}$ in the laboratory frame.}
\label{void}
\end{center}
\end{figure}
\begin{figure}[!h]
\begin{center}
\includegraphics[width=16.0cm]{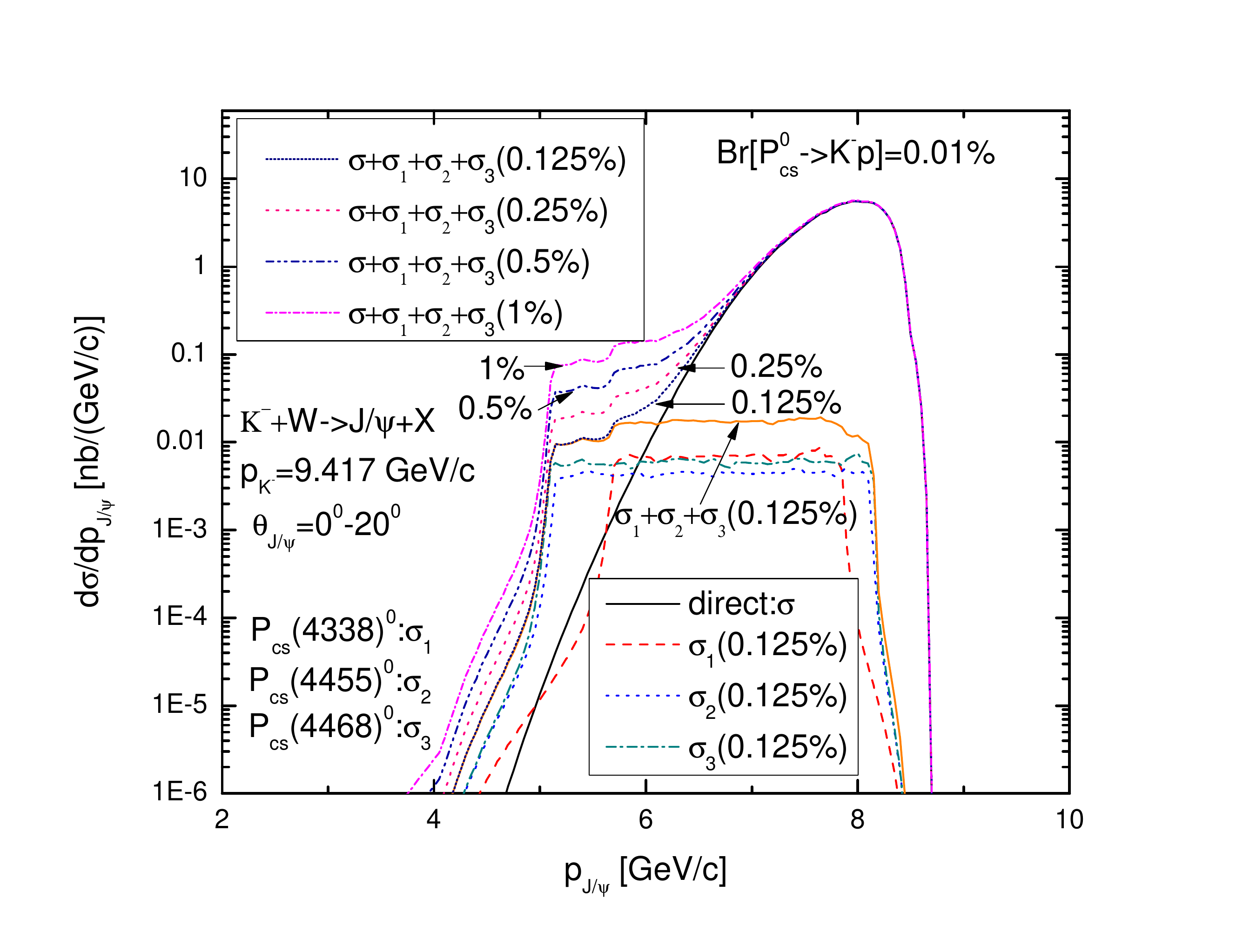}
\vspace*{-2mm} \caption{(Color online.) The direct non-resonant momentum distribution
of $J/\psi$ mesons, produced in the reaction ${K^-}{\rm ^{184}W} \to {J/\psi}X$
in the laboratory polar angular range of 0$^{\circ}$--20$^{\circ}$ and calculated in line with Eq. (28)
at initial antikaon resonant momentum of 9.417 GeV/c in the laboratory system (black solid curve).
The resonant momentum distributions of $J/\psi$ mesons, produced
in the two-step processes ${K^-}p \to P_{cs}(4338)^0 \to {J/\psi}\Lambda$,
${K^-}p \to P_{cs}(4455)^0 \to {J/\psi}\Lambda$ and
${K^-}p \to P_{cs}(4468)^0 \to {J/\psi}\Lambda$ and
calculated in line with Eq. (37) at the same incident antikaon momentum of 9.417 GeV/c
assuming that the resonances $P_{cs}(4338)^0$, $P_{cs}(4455)^0$
and $P_{cs}(4468)^0$ with the spin-parity assignments
$J^P=(1/2)^-$, $J^P=(1/2)^-$ and $J^P=(3/2)^-$, correspondingly,
all decay to the $K^-p$ and ${J/\psi}\Lambda$ modes
with branching fractions 0.01\% and 0.125\% (respectively, red dashed, blue dotted, dark cyan dashed-doted curves)
and their incoherent sum (orange solid curve).
Incoherent sum of the direct non-resonant $J/\psi$ momentum distribution and resonant ones,
calculated supposing that the resonances
$P_{cs}(4338)^0$, $P_{cs}(4455)^0$, $P_{cs}(4468)^0$
with the same spin-parity combinations all decay to the $K^-p$ and ${J/\psi}\Lambda$ with branching
fractions 0.01\% and 0.125, 0.25, 0.5 and 1\% (respectively, navy short-dotted, pink dotted, royal dashed-dotted-dotted and magenta short-dashed-dotted curves), all as functions of the $J/\psi$ momentum $p_{J/\psi}$ in the laboratory frame.}
\label{void}
\end{center}
\end{figure}
\begin{figure}[!h]
\begin{center}
\includegraphics[width=16.0cm]{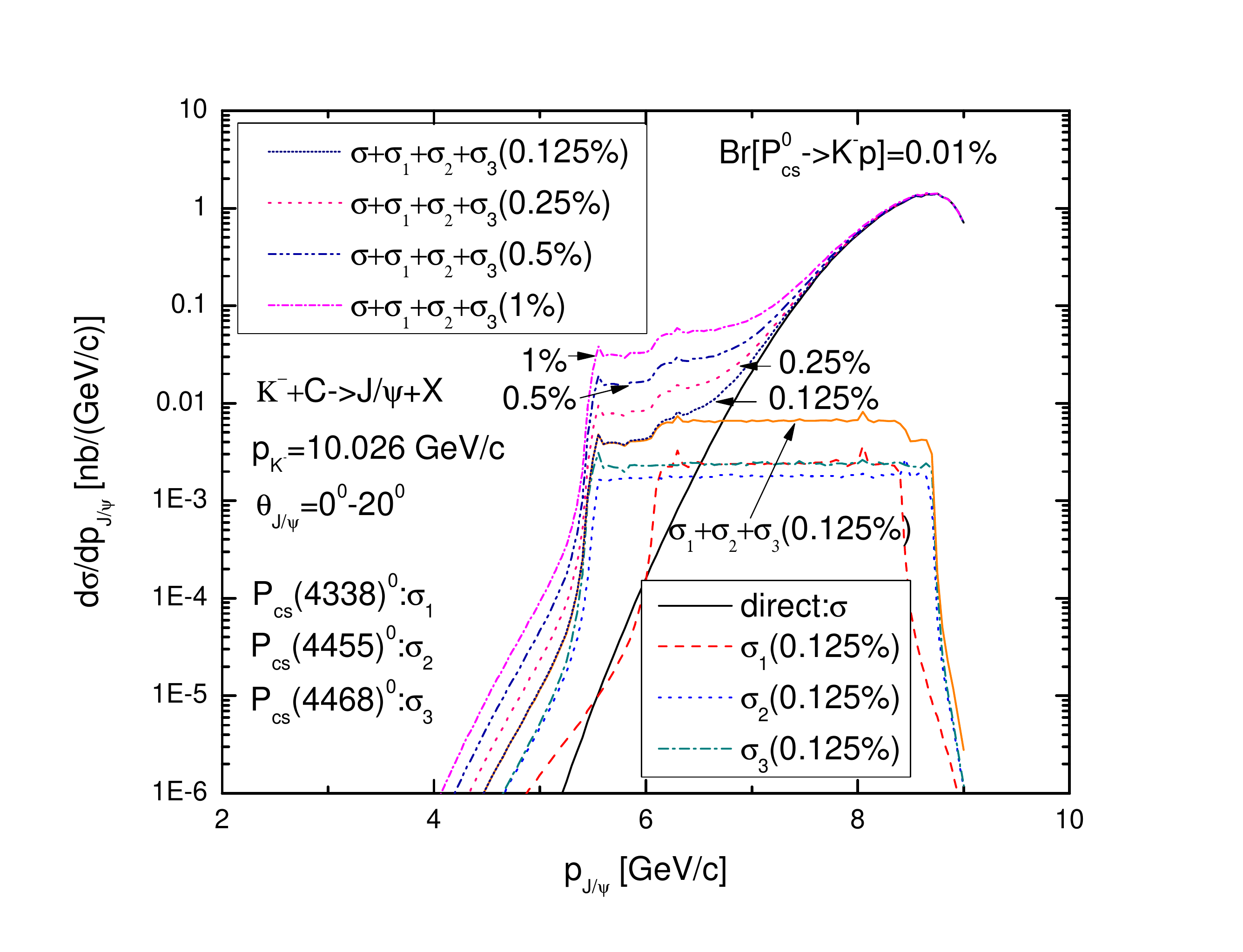}
\vspace*{-2mm} \caption{(Color online.) The same as in Fig. 9, but calculated for the initial antikaon momentum of 10.026 GeV/c.}
\label{void}
\end{center}
\end{figure}
\begin{figure}[!h]
\begin{center}
\includegraphics[width=16.0cm]{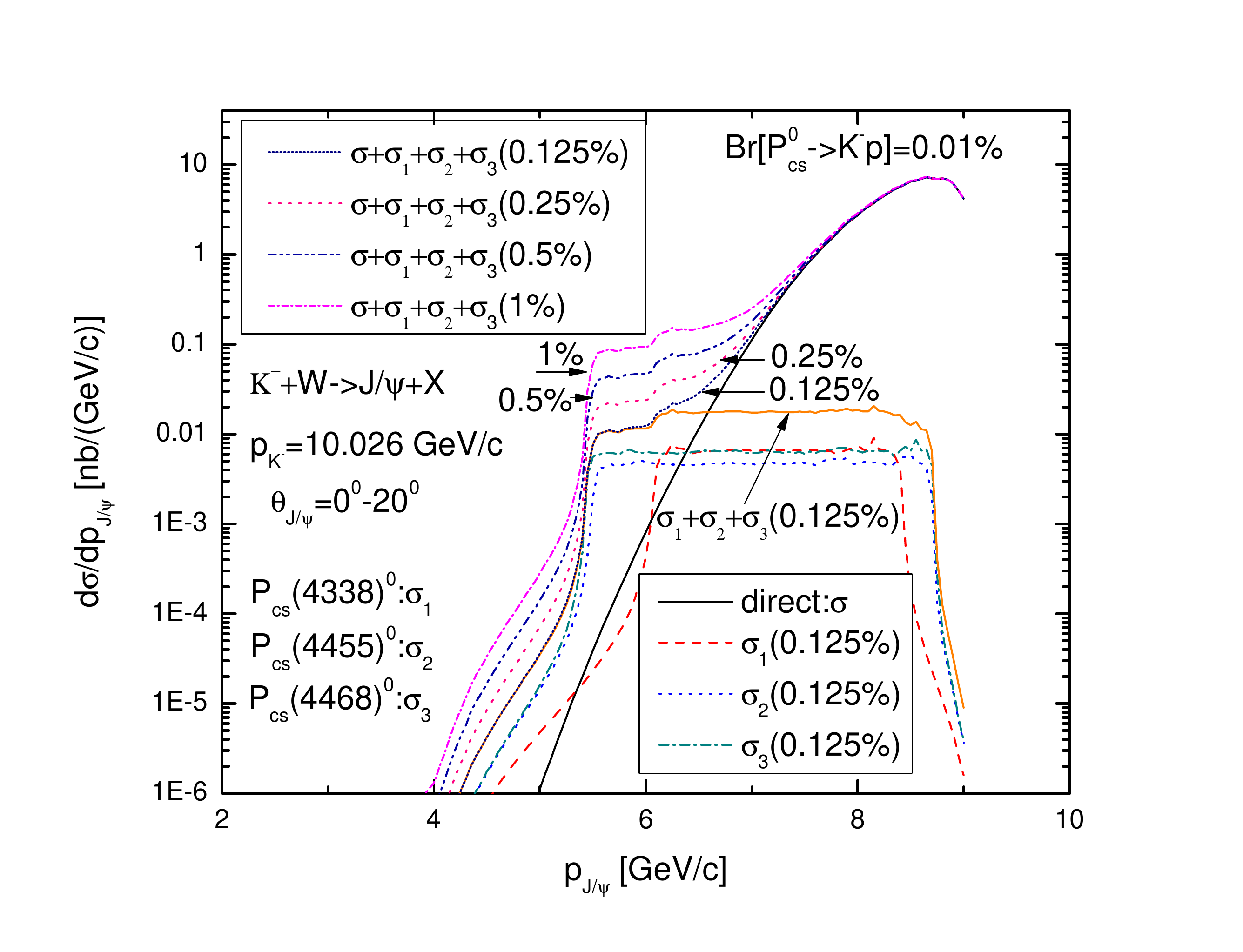}
\vspace*{-2mm} \caption{(Color online.) The same as in Fig. 10, but calculated for the initial antikaon momentum of 10.026 GeV/c.}
\label{void}
\end{center}
\end{figure}
\begin{figure}[!h]
\begin{center}
\includegraphics[width=16.0cm]{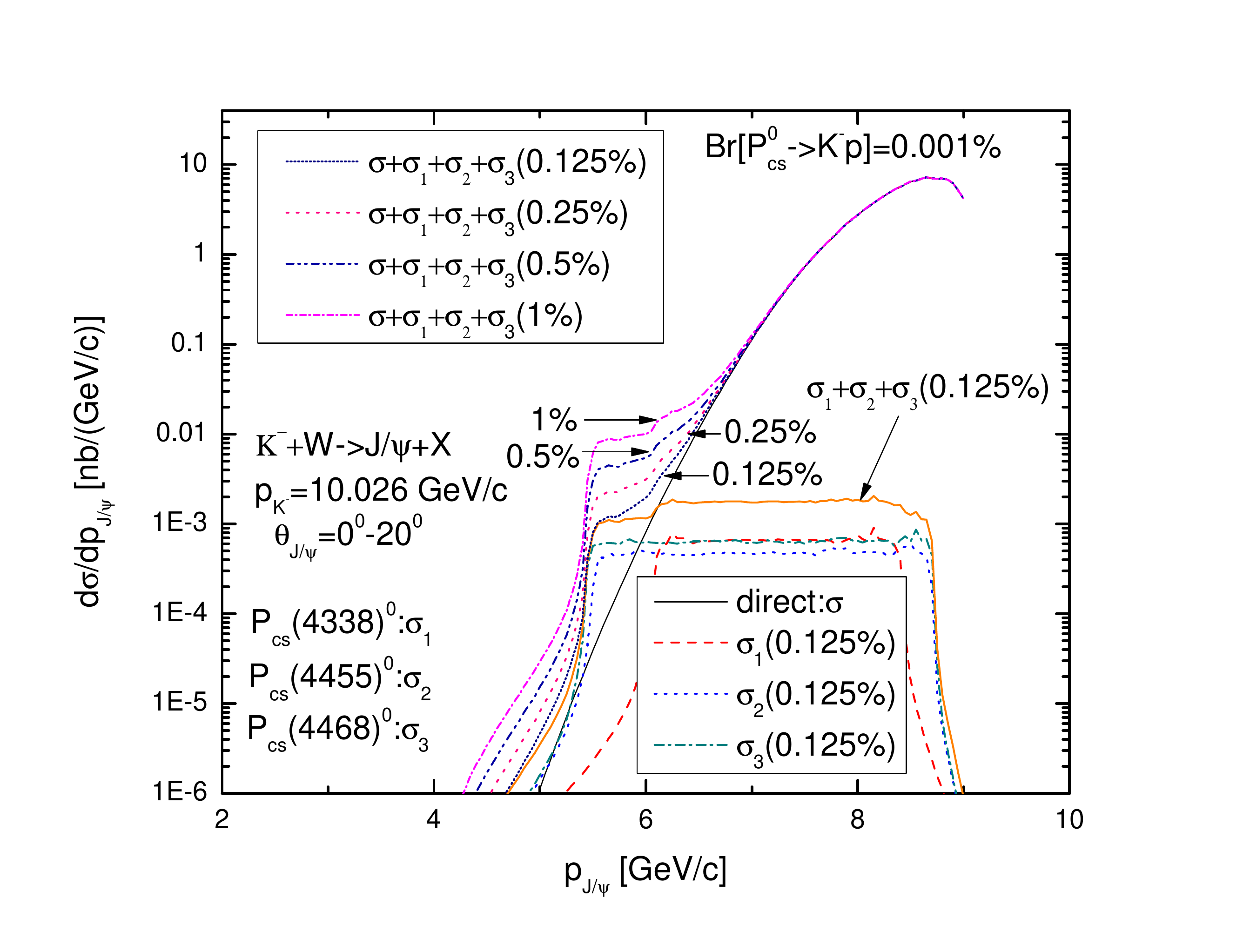}
\vspace*{-2mm} \caption{(Color online.) The direct non-resonant momentum distribution
of $J/\psi$ mesons, produced in the reaction ${K^-}{\rm ^{184}W} \to {J/\psi}X$
in the laboratory polar angular range of 0$^{\circ}$--20$^{\circ}$ and calculated in line with Eq. (28)
at initial antikaon resonant momentum of 10.026 GeV/c in the laboratory system (black solid curve).
The resonant momentum distributions of $J/\psi$ mesons, produced
in the two-step processes ${K^-}p \to P_{cs}(4338)^0 \to {J/\psi}\Lambda$,
${K^-}p \to P_{cs}(4455)^0 \to {J/\psi}\Lambda$ and
${K^-}p \to P_{cs}(4468)^0 \to {J/\psi}\Lambda$ and
calculated in line with Eq. (37) at the same incident antikaon momentum of 10.026 GeV/c
assuming that the resonances $P_{cs}(4338)^0$, $P_{cs}(4455)^0$
and $P_{cs}(4468)^0$ with the spin-parity assignments
$J^P=(1/2)^-$, $J^P=(1/2)^-$ and $J^P=(3/2)^-$, correspondingly,
all decay to the $K^-p$ and ${J/\psi}\Lambda$ modes
with branching fractions 0.001\% and 0.125\% (respectively, red dashed, blue dotted, dark cyan dashed-doted curves)
and their incoherent sum (orange solid curve).
Incoherent sum of the direct non-resonant $J/\psi$ momentum distribution and resonant ones,
calculated supposing that the resonances
$P_{cs}(4338)^0$, $P_{cs}(4455)^0$, $P_{cs}(4468)^0$
with the same spin-parity combinations all decay to the $K^-p$ and ${J/\psi}\Lambda$ with branching
fractions 0.001\% and 0.125, 0.25, 0.5 and 1\% (respectively, navy short-dotted, pink dotted, royal dashed-dotted-dotted and magenta short-dashed-dotted curves), all as functions of the $J/\psi$ momentum $p_{J/\psi}$ in the laboratory frame.}
\label{void}
\end{center}
\end{figure}

\section*{3. Results and discussion}

  The direct non-resonant $J/\psi$ production total cross section (24) on the target proton at rest
(black solid curves), the total cross section for the resonant $J/\psi$ production in the processes (30)/(31)
calculated on the basis of Eq. (32) for the considered spin-parity assignments of the $P_{csi}^0$ resonances
($i=$1, 2, 3) as well as for the branching ratios $Br[P_{csi}^0 \to {K^-}p]=$ 0.01\% and
$Br[P_{csi}^0 \to {J/\psi}\Lambda]=$ 0.125, 0.25, 0.5 and 1\% for all three
$P_{csi}^0$ states (red short-dashed curves) and the combined (non-resonant plus resonant)
$J/\psi$ production total cross section (blue dotted curves) are presented in Fig. 1 as functions of the $K^-$
meson laboratory momentum. The same as that shown in Fig. 1, but determined for the three branching ratios
$Br[P_{csi}^0 \to {K^-}p]=$ 0.001\%, is presented in Fig. 2 to see the sensitivity of the combined
$J/\psi$ production total cross section to these ratios.
One can see from these figures that the $P_{cs}(4338)^0$ state as well as $P_{cs}(4455)^0$
and $P_{cs}(4468)^0$ resonances manifest itself, respectively, as distinct narrow single peak as well as
two narrow overlapping peaks at antikaon resonant momenta $p^{\rm R1}_{K^-}=9.417$ GeV/c
as well as at $p^{\rm R2}_{K^-}=9.965$ GeV/c and $p^{\rm R3}_{K^-}=10.026$ GeV/c in the combined cross section,
only if $Br[P_{csi}^0 \to {K^-}p]=$ 0.01\% and $Br[P_{csi}^0 \to {J/\psi}\Lambda]=$ 1\% for all three
$P_{csi}^0$ states. In this case, in the resonance regions the resonant contributions are much larger than the
non-resonant ones of about 0.5 nb and the peak values of the strengths of these three peaks reach a well measurable values $\sim$ 2--3 nb. Hence, the background reaction $K^-p \to {J/\psi}\Lambda$ will
not influence the direct observation of the hidden-charm strange pentaquarks $P_{cs}(4338)^0$, $P_{cs}(4455)^0$
and $P_{cs}(4468)^0$ production at these momenta and in this case. If $Br[P_{csi}^0 \to {J/\psi}\Lambda] \le$ 0.5\%,
then the resonant $J/\psi$ yields are comparable to (or are less than) the non-resonant ones in the resonance
regions. As a result, the $P_{cs}(4338)^0$, $P_{cs}(4455)^0$ and $P_{cs}(4468)^0$ states do not appear as 
pronounced peaks in the combined cross section. This implies that will be hard to measure in this case the $P_{csi}^0$ pentaquark states in $J/\psi$ total production cross section on a proton target in the antikaon-induced reactions.
From Fig. 2 we see that the resonant $J/\psi$ production cross section is small compared to the non-resonant
contribution at all antikaon momenta considered for four adopted values for the three branching ratios
$Br[P_{csi}^0 \to {J/\psi}\Lambda]$, when $Br[P_{csi}^0 \to {K^-}p]=$ 0.001\%.
In this case, the combined total cross section of the reaction ${K^-}p \to {J/\psi}\Lambda$ has no distinct peak structures, corresponding to the $P_{csi}^0$ states, and it is practically not distinguished from that for
the background reaction.
In view of the above, it is natural to expect that the $P_{csi}^0$ signals could be well distinguished from the background reaction via the detailed scan of the $J/\psi$ total production cross section on a proton target in the
near-threshold $K^-$ momentum regions around resonant momenta of 9.417, 9.965 and 10.026 GeV/c in the future dedicated
experiment at J-PARC facility, if branching ratios $Br[P_{csi}^0 \to {K^-}p]$ $\sim$ 0.01\%
and $Br[P_{csi}^0 \to {J/\psi}\Lambda]$ $\sim$ 1\%.
To see experimentally such signals in the combined total cross section of the reaction
${K^-}p \to {J/\psi}\Lambda$, it is enough to have the $K^-$ momentum resolution (and the momentum binning)
of the order of 5 MeV/c. Thus, the c.m. energy ranges
$M_{csi}-{\Gamma_{csi}}/2 < \sqrt{s} < M_{csi}+{\Gamma_{csi}}/2$ ($i=$1, 2, 3) correspond to the laboratory
antikaon momentum regions of 9.401 GeV/c $< p_{K^-} <$ 9.433 GeV/c, 9.947 GeV/c $< p_{K^-} <$ 9.982 GeV/c
and 10.013 GeV/c $< p_{K^-} <$ 10.038 GeV/c, i.e. ${\Delta}p_{K^-}=$ 32, 35
and 25 MeV/c for the $P_{cs}(4338)^0$, $P_{cs}(4455)^0$ and $P_{cs}(4468)^0$, respectively.
This means that to resolve the peaks in the upper left panel of Fig. 1 the $K^-$ beam momentum resolution
(and the momentum bin size) of the order of 5 MeV/c are required. One may hope that this requirement will be
satisfied at the K10 beam line at the J-PARC facility [72]. To further motivate the conducting of the relevant
experiment at this facility, it is important to estimate the expected yields of the $P_{csi}^0$ signals from
the reactions $K^-p \to P_{csi}^0 \to {J/\psi}\Lambda$, $J/\psi \to e^+e^-$, $\Lambda \to p{\pi^-}$ at least
in the case, corresponding to the upper left panel of Fig. 1. Assuming incident antikaon beam intensity
$\sim$ 10$^6$ $K^-$/s in the peak antikaon momenta and 4g/cm$^2$ (57 cm) for the thickness of the liquid
hydrogen target [72], one could reach an integrated luminosity of 76 pb$^{-1}$ for a year of data taking.
For the pentaquark yield (for the total number of the $P_{csi}^0$ events) estimates in a one-year beam time,
one needs to multiply the above luminosity by the $P_{csi}^0$ production cross sections of 1.2, 1.2 and 2.3 nb
at resonant $K^-$ momenta of 9.417, 9.965 and 10.026 GeV/c as well as by the detection efficiency and by the
appropriate branching ratios $Br[{J/\psi} \to e^+e^-]$ $\approx$ 6\% and $Br[\Lambda \to p{\pi^-}]$ $\approx$ 64\%.
Even with a relatively low (and realistic) 10\% detection efficiency, we estimate about of 350, 350 and 670 events
per year for the $P_{cs}(4338)^0$, $P_{cs}(4455)^0$ and $P_{cs}(4468)^0$ signals, respectively.
For elastic (background) $J/\psi{\Lambda}$ production in the reaction $K^-p \to {J/\psi}\Lambda$ we have
$\sigma=0.54$ nb at considered $K^-$ momenta. This leads to $\approx$ 160 events per year. We see that
the elastic background is sufficiently small compared to the resonant yields.
Therefore, highly intense, energetic and resolution $K^-$ beam, which will be
available at this beam line, should allow to get a definite result for or against the existence of the genuine
$P_{cs}(4338)^0$ pentaquark state in the nature and the presence of two resonances in the peak of the
${J/\psi}\Lambda$ event distribution, associated to the $P_{cs}(4459)^0$, through the
scan of the $J/\psi$ total production cross section on a proton target in the resonance regions,
only if branching ratios $Br[P_{csi}^0 \to {K^-}p]$ $\sim$ 0.01\% and $Br[P_{csi}^0 \to {J/\psi}\Lambda]$ $\sim$ 1\%.
It is obvious that to see the $P_{csi}^0$ pentaquarks  experimentally in the cases, when more "softer" constraints
than those given above are imposed on the branching ratios $Br[P_{csi}^0 \to {K^-}p]$ and $Br[P_{csi}^0 \to {J/\psi}\Lambda]$, one needs to consider such observable, which is appreciably
sensitive to the $P_{cs}^0$ signal in some region of the available phase space.
For instance, the large $t$ region of the differential cross section $d\sigma/dt$
in the $J/\psi$-007 experiment [77], where the $t$-dependence of the background $J/\psi$ meson production
is suppressed while its resonant production is rather flat.
This is also supported by the findings of Refs. [81--83].

  Accounting for the above-mentioned, we focus now on the $J/\psi$ energy distribution
from the considered ${K^-}p \to {J/\psi}\Lambda$ elementary reaction.
Our model allows one to calculate the direct non-resonant $J/\psi$ energy distribution
from this reaction, the resonant ones from the production/decay sequences (30)/(31),
proceeding on the free target proton being at rest.
They were calculated in line with Eqs. (23), (36), respectively, for incident
antikaon resonant momenta of 9.417, 9.965 and 10.026 GeV/c.
The resonant $J/\psi$ energy distributions were determined for the employed spin-parity assignments
of the $P_{cs}(4338)^0$, $P_{cs}(4455)^0$, $P_{cs}(4468)^0$ resonances for branching fractions
$Br[P_{csi}^0 \to {K^-}p]=$ 0.01\% and $Br[P_{csi}^0 \to {J/\psi}\Lambda]=$~0.125\% for all three states.
These dependencies, together with the incoherent sum of the non-resonant $J/\psi$ energy distribution
and resonant ones, calculated supposing that all the resonances
$P_{cs}(4338)^0$ and $P_{csi}^0$ ($i=1$, 2, 3), $P_{cs}(4455)^0$ and $P_{csi}^0$ ($i=1$, 2, 3),
$P_{cs}(4468)^0$ and $P_{csi}^0$ ($i=1$, 2, 3) decay to the ${J/\psi}\Lambda$ final state
with four adopted options for the branching ratios $Br[P_{csi}^0 \to {J/\psi}\Lambda]$,
as functions of the $J/\psi$ total energy $E_{J/\psi}$ are shown, respectively, in Figs. 3, 4, 5.
The same as that shown in these figures, but calculated for the three branching ratios
$Br[P_{csi}^0 \to {K^-}p]=$ 0.001\%, is presented, respectively, in Figs. 6, 7, 8 to see
the sensitivity of the combined $J/\psi$ energy distribution to these ratios.
It can be seen from all these figures that while the resonant $J/\psi$ production differential
cross sections show a quite flat behavior at all allowed energies $E_{J/\psi}$,
the non-resonant cross section drops quickly as the energy $E_{J/\psi}$ decreases.
At incident $K^-$ meson resonant momenta of 9.417, 9.965 and 10.026 GeV/c of our interest and in the case when
$Br[P_{csi}^0 \to {K^-}p]=$ 0.01\% ($i=$ 1, 2, 3),
its strength is essentially larger than those of the resonant $J/\psi$ production cross sections,
calculated for the value of the branching ratios $Br[P_{csi}^0 \to {J/\psi}\Lambda]=$~0.125\%,
for "high" allowed $J/\psi$ total energies greater, correspondingly, than $\approx$~7.75, 8.25 and 8.5 GeV.
Whereas at "low" $J/\psi$ total energies (below, respectively, 7.75, 8.25 and 8.5 GeV)
the contribution from the resonance, decaying to the ${J/\psi}\Lambda$ mode with the branching ratio of 0.125\%,
with the centroid, correspondingly, at momenta of 9.417, 9.965 and 10.026 GeV/c, is much larger than the non-resonant one. When $Br[P_{csi}^0 \to {K^-}p]=$ 0.001\%, analogous $J/\psi$ total energies are somewhat smaller,
and they, respectively, are $\approx$~7.25, 7.75 and 8.0 GeV.
Thus, for example, in the first and second cases (what concerns the branching ratios $Br[P_{csi}^0 \to {K^-}p]$
considered), for the $J/\psi$ mesons with total energies about of 7.0 and 6.5 GeV
their resonant production cross section, calculated for the branching fraction of 0.125\% of its decay to the
${J/\psi}\Lambda$, is enhanced compared to the non-resonant one by sizeable
factors about of one, two and three order of magnitude at incident antikaon momenta of
9.417, 9.965 and 10.026 GeV/c, respectively. Furthermore, this contribution is also essentially larger
than those, arising from the decays of another two pentaquarks to the ${J/\psi}\Lambda$ with the
branching ratios $Br[P_{csi}^0 \to {J/\psi}\Lambda]=$~0.125\%, at the aforementioned "low" $J/\psi$ total energies.
As a consequence, for each $K^-$ beam momentum considered the $J/\psi$ meson combined energy distribution,
appearing from the direct $J/\psi$ meson production and from the decay of the pentaquark resonance located at this momentum to the ${J/\psi}\Lambda$ channel, reveals here a clear sensitivity to the employed  variations in the branching ratio of this decay.
Thus, for instance, for the $J/\psi$ mesons with total energy of 6.5 GeV and for the lowest incident antikaon
momentum of 9.417 GeV/c this $J/\psi$ combined distribution is enhanced
for the values of this ratio of 0.125, 0.25, 0.5, 1\% by substantial
factors of about 10.0, 20.0, 37.5, 80.0, respectively, as compared to that from the directly produced $J/\psi$
mesons in the case when $Br[P_{csi}^0 \to {K^-}p]=$~0.001\%. And for the highest initial $K^-$ meson momentum
of 10.026 GeV/c of our interest, at which the resonance $P_{cs}(4468)^0$ appears as a weak peak structure in the total
cross section of the exclusive reaction ${K^-}p \to {J/\psi}\Lambda$ only if $Br[P_{csi}^0 \to {J/\psi}\Lambda]=$~1\%,
the analogous factors become much larger and they are of about 0.25$\cdot$10$^3$, 0.5$\cdot$10$^3$,
1$\cdot$10$^3$, 2$\cdot$10$^3$, respectively. Similarly, for the $J/\psi$ mesons with total energy of 7.0 GeV,
for incident $K^-$ meson momenta of 9.417 and 10.026 GeV/c these factors are of about 13, 25, 50, 100 and
0.35$\cdot$10$^3$, 0.7$\cdot$10$^3$, 1.3$\cdot$10$^3$, 2.6$\cdot$10$^3$, correspondingly, if
$Br[P_{csi}^0 \to {K^-}p]=$~0.01\%.
Moreover, what is important, one can see that the above "partial"
combined energy distribution of the $J/\psi$ mesons, calculated at given values of the branching ratios
$Br[P_{csi}^0 \to {K^-}p]$ and $Br[P_{csi}^0 \to {J/\psi}\Lambda]$, is practically indistinguishable from their "total" combined energy distribution, arising from the direct and resonant $J/\psi$ meson production through all production/decay sequences (30)/(31), and determined at the same values of these branching ratios.
This means that the differences between the combined results, obtained by using a value of the branching fractions of the decays $P_{csi}^0 \to {J/\psi}\Lambda$ of 0.125\% and the non-resonant background,
as well as between the combined results, determined by employing the values of the branching ratios of these decays
of 0.125 and 0.25\%, 0.25 and 0.5\%, 0.5 and 1\%, are quite large and experimentally measurable at "low" $J/\psi$ total energies and for both adopted options for the branching fractions $Br[P_{csi}^0 \to {K^-}p]$ ($i=$~1, 2, 3).
Furthermore, for each initial antikaon resonant momentum considered
the observation here of the specific hidden-charm LHCb pentaquark with strangeness will be practically
not influenced by the presence of the another two strange hidden-charm pentaquark states and by the background reaction.
Since the $J/\psi$ production differential cross sections have a small absolute values $\sim$ 0.01--0.1 nb/GeV
and $\sim$ 0.1--1.0 nb/GeV at "low" $J/\psi$ total energies $E_{J/\psi}$ for branching fractions
$Br[P_{csi}^0 \to {K^-}p]=$~0.001 and 0.01\%, respectively, their measurement requires both high luminosities and large-acceptance detectors. Such measurement might be performed in the near future at the J-PARC facility
within the planned here experiments at the K10 beam line [72]. Accounting for the estimates of the $P_{csi}^0$
($i=$1, 2, 3) yields given before and the above-mentioned $J/\psi$ production differential cross sections at
"low" $J/\psi$ total energies, we can evaluate the numbers of events expected in this measurement in the $J/\psi$
energy range of 6.5--7.5 GeV for the $K^-p \to P_{csi}^0 \to {J/\psi}\Lambda$, $J/\psi \to e^+e^-$, $\Lambda \to p{\pi^-}$ reactions. They are about of 2.9--29 and 29--290 per year for $Br[P_{csi}^0 \to {K^-}p]=$~0.001 and 0.01\%, respectively. The number of background events is lower by several orders of magnitude. This means that the observation
of the hidden-charm strange pentaquarks $P_{csi}^0$ via the low-energy $J/\psi$ meson production on proton target
is quite optimistic at the J-PARC at least in the case when $Br[P_{csi}^0 \to {K^-}p]=$~0.01\% and $Br[P_{csi}^0 \to {J/\psi}\Lambda]=$~0.125, 0.25, 0.5 and 1\%. It should be pointed out that most likely it would be difficult to
determine the spin-parity quantum numbers of the $P_{csi}^0$ states by using the combined total and integrated over
production angles differential cross sections considered incoherently in the present work since that observables,
based on the squared absolute values of the resonant and non-resonant production amplitudes, depend weakly on them.
Such consideration is justified because the role of the interference effects between resonant and non-resonant
contributions in the $K^-p \to {J/\psi}\Lambda$ reaction as well as between different resonance states in the
observables considered is expected to be insignificant due to the fact that the $t$-channel $J/\psi$ production
contributes only to the forward angles in the c.m. frame while the $s$-channel resonances contribute in a full
solid angle [84]. A way of determining the $P_{csi}^0$ quantum numbers should be based on the model, which takes
into account the interference effects via adding the resonant and non-resonant production amplitudes coherently.
The results of such procedure depend, in particular, on these numbers. To probe them, another observables like
angular distributions [69, 85] or polarization observables (single and double) [86, 87], accounting for the spin correlations between the polarized target proton and the polarized recoil $J/\psi$ meson and $\Lambda$ hyperon
(spin of the incident $K^-$ meson is equal to zero), are needed. The consideration of this aspect is beyond the scope of the present paper. One may hope that it will be studied in the future work when the appropriate $K^-p$ data will appear to control that interference.

     The momentum dependencies of the absolute non-resonant, resonant and combined $J/\psi$ meson differential
cross sections, respectively, from the direct (5), two-step (30)/(31) and direct plus two-step $J/\psi$ production processes in $K^-$$^{12}$C and $K^-$$^{184}$W interactions, calculated on the basis of
Eqs. (28), (37) for laboratory polar angles of 0$^{\circ}$--20$^{\circ}$, for incident $K^-$ meson lowest resonant
momentum of 9.417 GeV/c and for branching fractions $Br[P_{csi}^0 \to {K^-}p]=$~0.01\%,
are shown, respectively, in Figs. 9 and 10. The same as in these figures, but determined for the initial highest antikaon resonant momentum of 10.026 GeV/c, is given in Figs. 11 and 12.
And finally, the same as that shown in Fig. 12, but calculated for the three branching ratios
$Br[P_{csi}^0 \to {K^-}p]=$ 0.001\%, is presented in Fig. 13 to see the sensitivity of the combined
$J/\psi$ momentum distribution to these ratios.
The resonant momentum distributions of the $J/\psi$ mesons in the two-step processes (30)/(31),
taking place on the bound protons of carbon and tungsten nuclei, were obtained
for four employed values of the branching ratios $Br[P_{csi}^0 \to {J/\psi}\Lambda]$.
It is seen from Figs. 9--12 that the total contribution to the $J/\psi$ production on both these nuclei, stemming
from all the intermediate $P_{csi}^0$ states, decaying to the ${J/\psi}\Lambda$
mode with branching fractions of 0.125\% (orange solid curves), shows practically flat behavior,
and it is significantly larger than that from the background process (5) (black solid curves)
in the "low"-momentum regions of 4.5--6.0 GeV/c and 4.5--6.5 GeV/c
for considered antikaon beam momenta of 9.417 and 10.026 GeV/c, respectively.
The results, presented in Fig. 13, show that such "low"-momentum region also exists and amounts to 4.5--6.0 GeV/c
for the initial $K^-$ momentum of 10.026 GeV/c and $Br[P_{csi}^0 \to {K^-}p]=$ 0.001\%. And in these
"low"-momentum regions, what is of primary importance, the decays $P_{cs}(4455)^0 \to {J/\psi}\Lambda$ and
$P_{cs}(4468)^0 \to {J/\psi}\Lambda$ are dominant.
As a consequence, the combined $J/\psi$ yield in them is practically completely governed by the presence of the $P_{cs}(4455)^0$ and $P_{cs}(4468)^0$ states in its production.
For given value of the branching ratios $Br[P_{csi}^0 \to {K^-}p]$,
the strength of the yield is almost completely determined by the branching ratios
$Br[P_{cs}(4455/4468)^0 \to {J/\psi}\Lambda]$, used in the calculations, with a value, which, on the one hand, depends weakly on the antikaon momentum and which, on the other hand, increases by a factor of about four for both antikaon beam momenta considered, when going from carbon target nucleus to tungsten one.
The value is still large enough to be measured, as one may hope, in the future experiment at the J-PARC.
This results in the  well separated and experimentally distinguishable
differences between all combined calculations, corresponding to the employed options for these ratios,
for both target nuclei, for both antikaon momenta considered and for both adopted options for the branching fractions
$Br[P_{csi}^0 \to {K^-}p]$ ($i=$~1, 2, 3).
Therefore, the $J/\psi$ meson production differential cross section measurements
on light and especially on heavy nuclear targets in the above $J/\psi$ "low"-momentum regions
at antikaon momenta in the resonance regions will open an opportunity to confirm or refute the
$P_{cs}(4455)^0$ and $P_{cs}(4468)^0$ states and, if they will be confirmed, to determine their branching
ratios to the ${J/\psi}\Lambda$ -- at least to distinguish between realistic options of 0.125, 0.25, 0.5 and 1\%.

  Taking into account the above considerations, we can conclude that the near-threshold $J/\psi$
energy and momentum distribution measurements in antikaon-induced reactions both on protons
and on nuclear targets will provide further evidence for the existence of the hidden-charm strange pentaquark
$P_{cs}(4338)^0$, for the presence of two resonances in the peak of the
${J/\psi}\Lambda$ mass spectrum around 4459 MeV, and will shed light on their decay rates to the channel ${J/\psi}\Lambda$.

\section*{4. Conclusions}

Accounting for the LHCb observation that the reported hidden-charm strange pentaquark $P_{cs}(4459)^0$
can split into two substructures, $P_{cs}(4455)^0$ and $P_{cs}(4468)^0$, with a mass difference of 13 MeV
as well as the newly observed hidden-charm pentaquark resonance $P_{cs}(4338)^0$ with strangeness, in this
paper we have studied within the double-peak scenario for the $P_{cs}(4459)^0$ state
the near-threshold $J/\psi$ meson production from protons and nuclei
by considering incoherent direct non-resonant (${K^-}p \to {J/\psi}\Lambda$) and two-step resonant
(${K^-}p \to P_{csi}^0 \to {J/\psi}\Lambda$, $i=1$, 2, 3; $P_{cs1}^{0}=P_{cs}(4338)^0$,
$P_{cs2}^{0}=P_{cs}(4455)^0$, $P_{cs3}^{0}=P_{cs}(4468)^0$) charmonium production processes
with the main goal of clarifying the possibility to observe within this scenario both above two substructures
contributing to the $P_{cs}(4459)^0$ state and the $P_{cs}(4338)^0$ resonance in this production.
We have calculated the absolute excitation functions, energy and momentum distributions
for the non-resonant, resonant and for the combined (non-resonant plus resonant) production
of $J/\psi$ mesons on protons as well as, using the nuclear spectral function approach,
on carbon and tungsten target nuclei at near-threshold incident antikaon beam momenta by assuming
the spin-parity assignments of the hidden-charm resonances
$P_{cs}(4338)^0$, $P_{cs}(4455)^0$ and $P_{cs}(4468)^0$ as $J^P=(1/2)^-$, $J^P=(1/2)^-$
and $J^P=(3/2)^-$ within four different realistic choices for the branching ratios
of their decays to the ${J/\psi}\Lambda$ mode (0.125, 0.25, 0.5 and 1\%) as well as for two options
for the branching fraction of their decays to the $K^-p$ channel (0.01 and 0.001\%).
It was shown that when the latter fraction is assumed to be 0.01\% then
will be very hard to measure the $P_{csi}^0$ pentaquark states through the scan of the $J/\psi$
total production cross section on a proton target in the near-threshold momentum region around the resonant
antikaon momenta of 9.417, 9.965 and 10.026 GeV/c if the considered branching ratios of the ${J/\psi}\Lambda$
decay mode $\sim$ 0.5\% and less. It was also found that if the branching fraction of the $P_{csi}^0 \to {K^-}p$
decays is suggested to be 0.001\% then these hidden-charm pentaquark states will be not accessible in such measurements.
It was further demonstrated that at these $K^-$ meson beam momenta the $J/\psi$ energy and momentum combined distributions of interest reveal noticeable sensitivity to all the above scenarios, respectively, at "low" $J/\psi$ total energies and momenta, which means that they may be an important tool to provide further evidence for the existence of
the strange hidden-charm pentaquark resonances $P_{cs}(4338)^0$, $P_{cs}(4455)^0$, $P_{cs}(4468)^0$
and to get valuable information on their decay rates to the $K^-p$ initial and ${J/\psi}\Lambda$ final states.
The measurements of these distributions could be performed in the future at the J-PARC facility. The present
model's predictions can also serve as a good starting point in supporting of these measurements.

\end{document}